\documentclass[prl,aps,reprint,amsmath,amssymb,showpacs,superscriptaddress]{revtex4-1}
\usepackage{graphicx}
\usepackage{xcolor}
\usepackage{verbatim}
\usepackage{hyperref}
\usepackage{tikz}
\hypersetup{colorlinks=true,
linkcolor=blue,
citecolor=blue,
urlcolor=blue,
filecolor=blue
}

\begin{document}

\title{Proposal for a transmon-based quantum router}

\author{Arnau Sala} \email{salacadellans@gmail.com}
\affiliation{Lorentz Institute for Theoretical Physics, University of Leiden, 2333 CA Leiden, The Netherlands}
\affiliation{Kavli Institute of Nanoscience, Delft University of Technology, Lorentzweg 1. 2628 CJ Delft, The Netherlands}
\author{M. Blaauboer} \email{m.blaauboer@tudelft.nl}
\affiliation{Kavli Institute of Nanoscience, Delft University of Technology, Lorentzweg 1. 2628 CJ Delft, The Netherlands}

\date{\today}

\begin{abstract}
We propose an implementation of a quantum router for microwave photons in a superconducting qubit architecture consisting of a transmon qubit, SQUIDs and a nonlinear capacitor. We model and analyze the dynamics of operation of the quantum switch using quantum Langevin equations in a scattering approach and compute the photon reflection and transmission probabilities. For parameters corresponding to up-to-date experimental devices we predict successful operation of the router with probabilities above 94\%.
\end{abstract}

%
%

\maketitle

\section{Introduction}
Given recent advances in experimental quantum computation, where few-qubit systems have been realized~\cite{haffner,dicarlo,vandersar,lucero,monroe,barz}, a considerable amount of research is currently devoted to investigating larger-scale systems such as networks for quantum communication~\cite{stucki,ritter,garcia}. In such systems, where information must be coherently transported over long distances, photons are suitable candidates as quantum information carriers because of their long coherence times. Solid-state devices, on the other hand, seem preferable for storage of quantum information~\cite{chiorescu,wu}. Essential components in any quantum communication toolbox are, thus, devices capable of directing photons through different channels~\cite{ritter,garcia,lemr,lemr2,qu}. These devices include single-photon transistors~\cite{chang,neumeier,manzoni}, switches or routers~\cite{hoi,agarwal,li,lu,yan}, single-photon beam splitters~\cite{prl104,hoffmann}, etc.

So far, several approaches have been proposed~\cite{[{See }][{ and references therein.}]pra78} for building a quantum switch, such as an optical implementation using polarized photons and trapped atoms or a phase gate implementation~\cite{pra78}. Also, proposals for a beam splitter based on Superconducting Quantum Interference Devices (SQUIDs) to route photons~\cite{prb78,prl110,prl104,hoffmann,kyaw} or using toroidal resonators~\cite{pra86} have been put forward.

Here we propose a solid-state implementation of a quantum router using superconducting qubits. Given an $n$-photon input, consisting of a train of photon pulses where each of them can be in two states (or in a coherent superposition of both), the quantum switch absorbs the first photon and forwards the next $n-1$ photons to a path determined by the state of the first absorbed photon. Our proposed device can be integrated in a larger network, where the output of one router is the input of the next. As a result, since the routers are controlled by the input signals there is no need to control the network externally. Also it requires the same number of photons to send a signal through two different paths with the same number of nodes, while other proposals~\cite{hoi,garcia} may require less photons for some preferred paths, leading to a network where some routes (e.g., a route that at each bifurcation node takes the rightmost output path) have higher efficiency than others. We analyze the dynamic operation of the router using quantum Langevin equations combined with input/output scattering formalism, taking into account decoherence due to relaxation and dephasing. We predict successful operation of the router with probabilities above 94\% under realistic experimental conditions.

The paper is organized as follows. First we present our proposal for the superconducting circuit that operates as a quantum router and derive the effective Hamiltonian of the system. We then analyze the dynamics of operation of the quantum switch using a scattering approach and calculate the probabilities of reflection and transmission of an incoming photon. In the last section conclusions and a discussion of the possible applications of this quantum device are presented. \\

\section{Model}
\begin{figure}
\includegraphics[width=\linewidth]{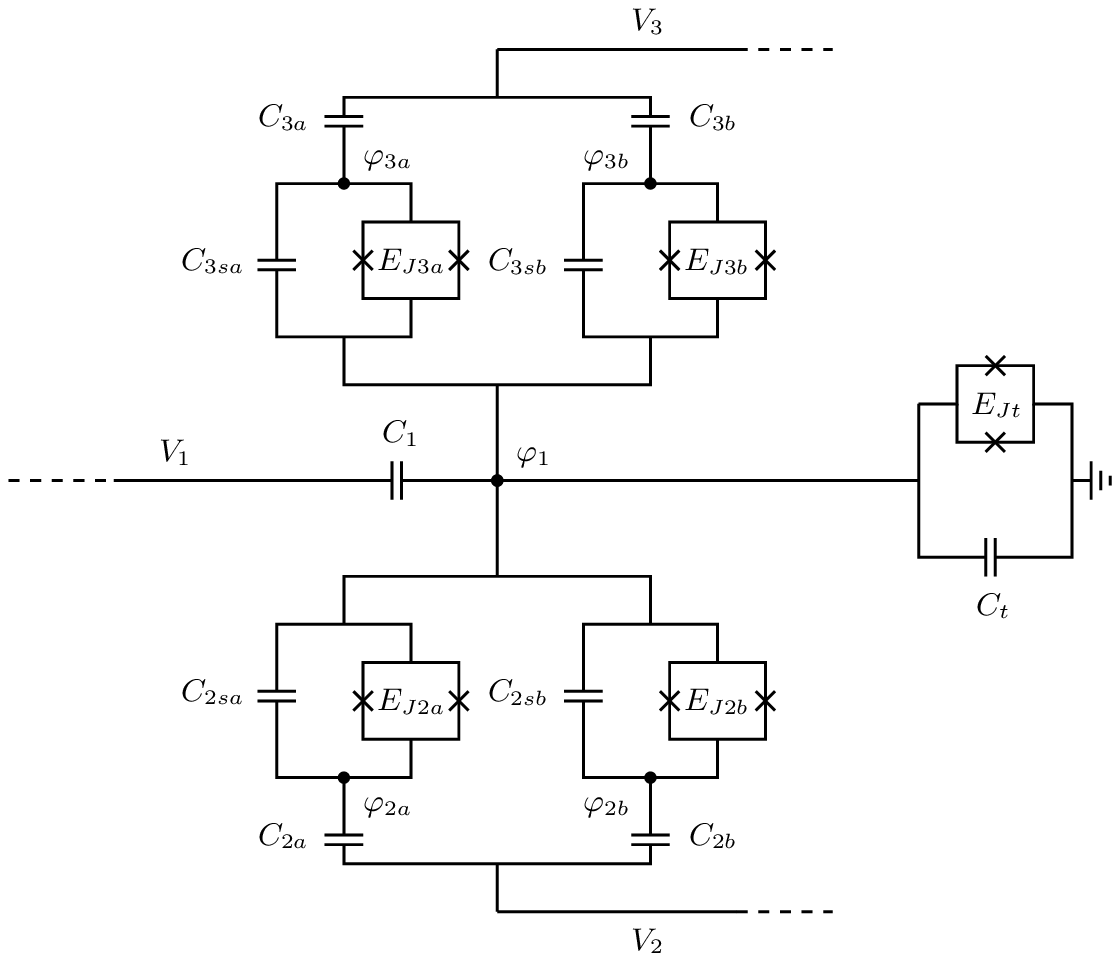}
\caption{Circuit QED device proposed to operate as a quantum router. This device is composed of four SQUIDs, each with a capacitor ($C_{2sa},~C_{2sb},~C_{3sa},~C_{3sb}$) and a pair of Josephson junctions (with energies $E_{J2a}$, etc.). It also contains a transmon qubit, consisting of a capacitor ($C_t$) and a pair of Josephson junctions ($E_{Jt}$). These five elements are capacitively coupled to an incoming transmission line ($V_1$) and to two outgoing transmission lines ($V_2$ and $V_3$). The fluxes $\varphi_i$, in the nodes of the circuit, are quantized variables (see \cite{Note1} for the quantization of these variables) and describe the absorption and emission of incoming photons by the transmon and SQUIDs. \label{fig:device}}
\end{figure}

A schematic of the device we propose, based on circuit quantum electrodynamics (cQED), is depicted in FIG.~\ref{fig:device}. The device is composed of three transmission lines capacitively coupled to four SQUIDs and a transmon, acting as artificial atoms that absorb and reflect or transmit the photons forward. The device operates as follows: the first photon of a register that arrives at the switch is absorbed by the transmon which, after being excited, modifies the energy spectrum of the SQUIDs in such a way that the next photons can only be absorbed by the two SQUIDs labeled with $2a$ or $2b$ ($3a$ or $3b$) if the transmon is in its first (higher) excited state.

The excited SQUID then decays while emitting a photon into its corresponding outgoing transmission line. The transmon also decays emitting a photon into the incoming transmission line. The capacitances $C_{2sa}$, $C_{2sb}$, $C_{3sa}$ and $C_{3sb}$ are chosen such that they are smaller than any other capacitance in the system. In this way, the transmission of the control photon, which is the element of the register that controls the operation of the router, is prevented~\cite{Note1}. Also the coupling strength between the SQUIDs and their farthest transmission line is strongly reduced if this condition on the capacitances is satisfied.

The dynamics of the router in FIG.~\ref{fig:device} is described by an effective Hamiltonian~\cite{Note1}
\begin{equation}\label{eq:heff}
H_{eff} = H_{sys} + H_c + H_T.
\end{equation}
Here $H_{sys}$ describes the energy levels of the transmon and SQUIDs and also the interaction between them:
\begin{align}\label{eq:hsys}
H_{sys} =&  \sum_{i=1}^3 \omega_{Ti} a^\dagger_{Ti} a_{Ti} + \sum_k \omega_{k} a^\dagger_{k} a_{k} \notag \\
&           - \sum_{i=1}^3 \sum_k J_{ik}a^\dagger_{Ti} a_{Ti}a^\dagger_{k} a_{k},
\end{align}
with $k \in \{ 2a, 2b, 3a, 3b \}$. Here $a^\dagger$ and $a$ denote the creation and annihilation operators for each of the energy levels of the system. These operators create an excitation with energy $\omega_{Ti}$ in the transmon or $\omega_k$ in the SQUIDs. The last term describes the density-density interaction between the $i$-th level of the transmon and the excitations in the $k$-th SQUID, with interaction strength $J_{ik}$.

$H_c$ describes the exchange interaction between the transmon and SQUIDs with the transmission lines, which are modeled as a bath of harmonic oscillators~\cite{houches,fan} with ladder operators $b_1$, $b_2$ and $b_3$ for the first, second and third transmission lines, respectively:
\begin{align}\label{eq:hc}
H_c =&      \int dp \left[ \frac{a^\dagger_{T1} b_1 (p)}{\sqrt{\pi \tau_{T1}}} + \frac{a^\dagger_{T3} b_1 (p)}{\sqrt{\pi \tau_{T3}}}\right.\nonumber \\
&           \qquad + \left( \sqrt{\frac{2}{\pi\tau_{T1}}} - \sqrt{\frac{3}{\pi\tau_{T3}}} \right) a^\dagger_{T2} a_{T1} b_1 (p) \nonumber \\
&           \qquad + \left( \frac{a^\dagger_{2a}}{\sqrt{\pi \tau_a}} + \frac{a^\dagger_{2b}}{\sqrt{\pi \tau_b}} \right) \left( b_1(p) + b_2(p) \right) \nonumber \\
&           \qquad \left. + \left( \frac{a^\dagger_{3a}}{\sqrt{\pi \tau_a}} + \frac{a^\dagger_{3b}}{\sqrt{\pi \tau_b}} \right) \left( b_1(p) + b_3(p) \right) + h.c. \right].
\end{align}
The interaction strength of this coupling is given by $\tau_{T1}$ and $\tau_{T3}$, which are the lifetimes of the excited levels of the transmon, and also $\tau_a$ and $\tau_b$, which are the lifetimes of the SQUIDs. Note that the transmon is only coupled to the incoming transmission line and that the second level of the transmon is not coupled (directly) to the transmission lines~\cite{Note2}. The coupling between the third level of the transmon and the transmission line ---which is not present in the Jaynes-Cummings Hamiltonian, the expression that usually describes cQED systems similar to FIG.~\ref{fig:device}~\cite{dicarlo,lucero,koch}--- is achieved by introducing a nonlinear capacitor in the transmon~\cite{Note3} (see also Sup. Mat.). This nonlinear capacitor can be realized, e.g., by placing carbon nanotubes between the plates of the capacitor~\cite{Ilani}. These give rise to an energy spectrum that has the form $E(V)= \frac{C}{2} (V^2 + \alpha V^4)$ for small voltages $V$\cite{Note4,Akinwande2}.

\begin{figure}
\includegraphics{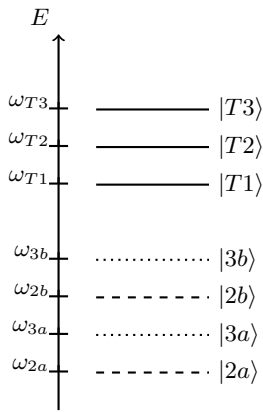}
\caption{Schematic representation of the energy spectrum of the Hamiltonian in Eq.~\eqref{eq:hsys}. This spectrum only contains the `single-photon levels', i.e., the levels that are accessible only by absorbing one single photon (plus $\left| T2 \right\rangle$, for completeness). Energy levels represented with solid lines are coupled only to the incoming transmission line. Dashed lines describe levels coupled to the second outgoing transmission line and also the incoming one. Dotted lines describe levels coupled to the third outgoing transmission line and also the incoming one. The separation of the energy levels is not to scale.\label{fig:spect1}}
\end{figure}
\begin{figure}
\includegraphics[scale=1]{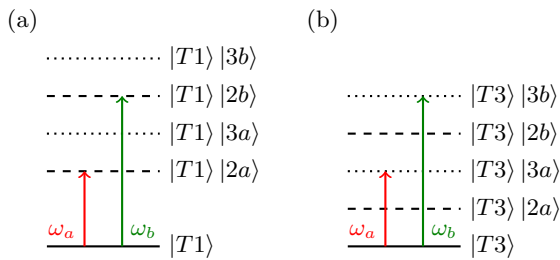}
\caption{(Color online) Energy spectrum of the Hamiltonian in Eq.~\eqref{eq:hsys} after the transmon has been excited. (a) After the first energy level of the transmon has been excited, the next photon, which has energy either $\omega_a$ or $\omega_b$ (or a superposition of both) can only excite the energy levels coupled to the second outgoing transmission line. (b) If the third energy level of the transmon has been excited, the same photon can only excite the energy levels coupled to the third outgoing transmission line.\label{fig:spect2}}
\end{figure}

The energy spectrum of the device, described by the Hamiltonians $H_{sys}$ and $H_c$ [Eq.~(\ref{eq:hsys}) and~ (\ref{eq:hc})], is shown in FIG.~\ref{fig:spect1}. In this figure the three levels of the transmon are represented with solid lines. The energy levels of the SQUIDs, in dashed lines (coupled to the second transmission line) and dotted lines (coupled to the third transmission line), do not coincide since all the capacitances and Josephson energies are different. After sending a control photon ---with energy $\omega_{T1}$ or $\omega_{T3}$--- into the router (see Fig.~\ref{fig:spect2}), due to the coupling between the transmon and SQUID energy levels described in Eq.~(\ref{eq:hsys}), the energy needed to create an excitation in SQUID $2a$ when the first level in the transmon is occupied can be made equal to the energy needed to create an excitation in SQUID $3a$ when the third level in the transmon is occupied (likewise for $2b$ and $3b$). This can be done by tuning the energies of the Josephson junctions of the system. Thus, the transmon ---the control element of the router--- forwards a photon with energy $\omega_a$ (or $\omega_b$) to the second or third transmission line, depending solely on the state of the control photon.

The transmission lines are described by a continuum of oscillating modes as~\cite{houches,fan}:
\begin{equation}\label{eq:ht}
H_T = \int dp\,p \left( b^\dagger_1(p) b_1(p) + b^\dagger_2(p) b_2(p) + b^\dagger_3(p) b_3(p) \right).
\end{equation}

\section{Analysis}
In order to analyze the dynamics of the quantum router we have studied the scattering of one and two photons by our proposed device. Here we calculate the probabilities of reflection and transmission through each of the output channels.

We start by deriving the Langevin equations from the effective Hamiltonian $H_{eff}$ [Eq.~(\ref{eq:heff})] including relaxation and dephasing~\cite{Note1,ithier}. We then obtain the equations of motion within the input/output formalism of quantum optics~\cite{neumeier,gardiner,fan}. From these equations we obtain the scattering amplitude of transmission and reflection of photons, assuming a Lorentzian pulse shape for the incoming photons~\cite{neumeier}.


Due to the multiple energy levels of the transmon (depicted in FIG.~\ref{fig:spect1}), the equation of motion for this element of the system ---in terms of the operators $a_1$ and $a^\dagger_1$ introduced after quantizing the fluxes~\cite{Note1}--- is a complicated equation to work with. In order to simplify the calculations we introduce three projection operators $a_{T1}$, $a_{T2}$ and $a_{T3}$ and their hermitian conjugate ---whose definition can be found in the Supp. Mat.--- to replace $a_1$ and $a^\dagger_1$. Moreover, we have limited our Hilbert space by considering only three excited energy levels in the transmon and only one in each of the SQUIDs. These operators, with the properties
\begin{eqnarray}
a^\dagger_{Ti} \left| GS \right\rangle =& \left| Ti \right\rangle \nonumber \\
a^\dagger_{Ti} \left| Tj \right\rangle =& 0 \qquad \forall~ i,j,
\end{eqnarray}
and also
\begin{eqnarray}
a_{Ti} \left| Ti \right\rangle =& \left| GS \right\rangle \nonumber \\
a_{Ti} \left| GS \right\rangle =& 0 \nonumber \\
a_{Ti} \left| Tj \right\rangle =& 0 \qquad \forall~ i \ne j,
\end{eqnarray}
with $\left| GS \right\rangle$ being the ground state of the transmon and $\left| Ti \right\rangle$ its excited states, are projection operators that do not satisfy the usual commutation relations, so special care has to be taken when working with them. On the other hand, one of the advantages of using these three operators is that, although we will have three equations for the transmon instead of one, they are easier to solve. The use of these three operators instead of the original one does not change the model as long as the number of excited states of the transmon is limited to three. Another advantage of using this notation is that the final quantized Hamiltonian takes a simpler form [the results being Eqns.~(\ref{eq:hsys}) and~(\ref{eq:hc})] and is easier to interpret: the operator describing the occupation of the $i$-th level of the transmon is simply $a^\dagger_{Ti}a_{Ti}$.

Before deriving the Langevin equations we must find an expression that describes the incoming and outgoing photons in the input/output formalism. Following~\cite{gardiner} we introduce the operators $b^\dagger_{2in}$ and $b^\dagger_{2out}$, and their hermitian conjugates, which create an incoming and an outgoing photon in the second transmission line (the same procedure must be applied to the other two transmission lines). These operators are defined as (see e.g.~\cite{neumeier,gardiner}):
\begin{equation}
b^\dagger_{2in/out}(t) = \frac{1}{\sqrt{2 \pi}} \int_{-\infty}^\infty d\omega\, e^{-i\omega(t-t_1)}b_2(\omega,t_1),
\end{equation}
where $\omega$ is the frequency of the photon and $b_2(\omega,t_1)$ is an initial (or final) value of the Heisenberg operator defined at a time $t_1 \to -\infty$ for the input operator (or $t_1 \to +\infty$ for the output operator). The Langevin equations of the `static' variables of the system in terms of $b_{1in}$, $b_{2in}$ and $b_{3in}$ are shown in the Supp. Mat. Once the Langevin equations are found, relaxation and dephasing ratios are then introduced into this set of equations by following Refs.~\cite{ithier,gardiner}.

Within the input/output formalism, the scattering amplitude for a photon with frequency $k$ in transmission line 1 to be reflected or transmitted in a transmission line $i$ with frequency $p$ is given by
\begin{equation}
S(p,k) = \left\langle 0 \right| b_{i,out}(p) b^\dagger_{1in}(k) \left| 0 \right\rangle.
\end{equation}
In the case that multiple photons are sent in, the scattering amplitude reads
\begin{multline}
S(p_1,\dots,p_n;k_1,\dots,k_n)      =\\
                                    = \left\langle 0 \right| b_{i,out}(p_1) \dots b_{i',out}(p_n) b^\dagger_{1in}(k_1) \dots b^\dagger_{1in}(k_n) \left| 0 \right\rangle.
\end{multline}
These scattering amplitudes are valid for incoming photons with a definite frequency $k_1,\dots,k_i$ each or a superposition of these. Assuming that the incoming photons have a frequency distribution given by a Lorentzian centered at $k_i=\omega_i$ and with width $1/\tau_i$, the scattering amplitude of a photon with frequency $\omega_1$ is given by
\begin{equation}
\beta(p) = \int \frac{dk_1}{\sqrt{\pi \tau_1}} \frac{1}{i(\omega_1 - k) + \frac{1}{\tau_1}} S(p_1,k_1).
\end{equation}
The probability of this process to happen is given by:
\begin{equation}
P = \int dp \left| \beta(p) \right|^2.
\end{equation}

Let us consider again the single photon processes. For a good performance of the quantum router, the first photon to arrive must be absorbed by the transmon and be reflected back after all the other photons have been transmitted. It is, thus, necessary to find a procedure to find out whether the photon has been reflected after being absorbed or without being absorbed. For this purpose we describe the incoming photons as even modes given by the superposition of left and right-moving modes as $b_{1in}(k) = (r_{1in}(k) + l_{1in}(k))/\sqrt{2}$~\cite{fan}. With this procedure one can see, after computing the scattering amplitudes and the probabilities, that a photon described as an even mode is absorbed (and reflected) with a large probability (see Tables~\ref{tab:1ph} and ~\ref{tab:2ph} below). Moreover, the probability of transmission of subsequent photons is enhanced if these are even modes.


The probabilities of reflection and transmission of photons are obtained by numerically integrating the squared scattering amplitude over all momenta and imposing the conditions $\omega_{2a}-J_{12a} = \omega_{3a}-J_{33a}=\omega_a$ and, similarly, $\omega_{2b}-J_{12b} = \omega_{3b}-J_{33b}=\omega_a$. This ensures that the router is controlled by the state of the first incoming photon.

The results of our calculations are shown in Tables~\ref{tab:1ph} and \ref{tab:2ph}. The first table contains the probabilities of transmission of an incoming photon with different frequencies $\omega_{T1}$, $\omega_{T3}$, $\omega_{2a}$ and $\omega_a$ through the second and third transmission lines together with the probabilities of reflection in case this photon is described as an even mode or as a right-moving mode (see Sup. Mat. and~\cite{fan} for a description of these modes). In the case of a photon with energy $\omega_{T1}$ or $\omega_{T3}$, since both energy levels are coupled only to the incoming transmission lines [see Eq.~(\ref{eq:hc})], we find a large probability of reflection, above 94\%. It is interesting to notice the different behavior of the router with respect to the form of the modes. In case the photons are described by even modes, the probability amplitude of reflection scales as
\begin{equation}
S_{even} \sim -\frac{1-\frac{3}{4} \gamma \tau_{T1}}{1+\frac{3}{4} \gamma \tau_{T1}}
\end{equation}
for a photon with frequency $\omega_{T1}$. In this expression, $\gamma$ is the rate of decoherence and $\tau_{T1}$ is the lifetime of the excited energy level in the transmon. In case of right-moving modes, since a right-moving mode is reflected as a left-moving mode if it is absorbed and as a right-moving mode if it is reflected but not absorbed~\cite{neumeier}, the corresponding expression in  Table~\ref{tab:1ph} represents the probability that a photon is absorbed and reflected, with an amplitude that is given by
\begin{equation}
S_{right} \sim -\frac{1}{1+\frac{3}{4} \gamma \tau_{T1}}.
\end{equation}

\begin{table}
\begin{ruledtabular}
\begin{tabular}{c c c c c}
       &    $\omega_{T1}$     &     $\omega_{T3}$     &     $\omega_{2a}$     &     $\omega_{a}$      \\ \hline
Refl. even modes  &     0.997 &     0.947             &     $1.9\cdot 10^{-2}$&     1                 \\
Refl. right-moving&     0.998 &     0.973             &     0.238             &     $1.6\cdot 10^{-5}$\\
Transmission 2    &     0     &     0                 &     0.952             &     $9.8\cdot 10^{-6}$\\
Transmission 3    &     0     &     0                 &     $1.6\cdot 10^{-6}$&     $2.3\cdot 10^{-6}$
\end{tabular}
\end{ruledtabular}
\caption{Probabilities of reflection and transmission of an incoming photon. The probabilities in this table have been computed for different frequencies of the incoming photons and considering both even modes and right-moving modes. Similar values can be obtained for left-moving modes. The probability of reflection of a right-moving photon, as computed here, is equivalent to the probability of absorption plus reflection of the incoming photon. \label{tab:1ph}}
\end{table}

A photon will only be transmitted if it can excite one of the SQUIDs. Table~\ref{tab:1ph} also shows that a photon with energy $\omega_{2a}$, which coincides with the energy of the excited state of one of the SQUIDs (see FIG.~\ref{fig:spect1}), is totally transmitted if it is described as an even mode and only partially transmitted if it is a right-moving mode. Since we are only interested in a scenario where the photons absorbed by the SQUIDs are always transmitted, we only consider even modes from now on. Notice that the probability of transmission of a photon indeed goes to one in case it is transmitted to the nearest transmission line of the excited SQUID (with even modes) and vanishes in case it is transmitted to the other outgoing transmission line.

More interesting is the case where two photons (a control and a target photons) are sent into the router. In this case, while the transmon is in the (slow) process of absorption and emission of the control photon, the second (target) photon is absorbed and emitted in a faster process by a SQUID before the transmon returns to its ground state. The data corresponding to these scattering processes is shown in Table~\ref{tab:2ph}. In this case, a photon described by an even mode with energy either $\omega_a$ or $\omega_b$ is transmitted to the second transmission line with a probability above 95\% if the first level of the transmon is excited. The probability amplitude of transmission of a photon with energy $\omega_a$ to the second outgoing transmission line is then given by
\begin{equation}
S_{2} \sim \frac{1}{1+\frac{3}{8} \gamma \tau_a}.
\end{equation}
If the third instead of the first level is excited, the photon is forwarded to the third outgoing transmission line with the same probabilities as in the case a photon is forwarded to the second transmission line when the first level of the transmon is excited. For any other frequency of the incoming photons these are not transmitted. \\

\begin{table}
\begin{ruledtabular}
\begin{tabular}{c c c c}
                        &     $\omega_{2a}$           &     $\omega_a$              &     $\omega_b$              \\ \hline
Transmission 2 (T1)     &     $4.45 \cdot 10^{-6}$    &     0.952                   &     0.964                   \\
Transmission 3 (T1)     &     $7.58 \cdot 10^{-5}$    &     $1.63 \cdot 10^{-6}$    &     $1.47 \cdot 10^{-6}$    \\
Transmission 2 (T3)     &     $2.43 \cdot 10^{-7}$    &     $7.12 \cdot 10^{-7}$    &     $2.94 \cdot 10^{-6}$    \\
Transmission 3 (T3)     &     $4.45 \cdot 10^{-6}$    &     0.952                   &     0.964
\end{tabular}
\end{ruledtabular}
\caption{Probabilities of reflection and transmission of a photon while one of the levels ---$T1$ or $T3$--- of the transmon is excited. Only even modes have been considered (see the text).\label{tab:2ph}}
\end{table}

\section{Conclusions and Outlook}
We have proposed and analyzed a transmon-based quantum router containing a nonlinear capacitor that operates with photons in the microwave regime. The nonlinear capacitor is the element responsible for the photon-transmon interaction. We predict successful operation of the quantum router with probabilities above 94\% for current experimental parameters. That is, the probability that a photon is transmitted successfully to the target channel, assuming that the transmon has been excited with a high probability (as the calculations suggest) and that the photon is transmitted before the transmon returns to its ground state, is above 94\%. This router can be built with a set of capacitors and Josephson junctions (see Ref.~\cite{Note1} for a proposal of such parameters) in which two of the SQUIDs (coupled to the same outgoing transmission line) can be excited by a photon with frequency $\omega_a$ or $\omega_b$ if one (or the other) level of the transmon is excited. 

The quantum router can be used as a single-photon transistor, with the distinctive characteristics that both the control and target photons can come from the same transmission line (i.e., from the same source) and that the target photon can be transmitted into two different transmission lines. The control photon thus not only controls whether the target photon is transmitted or not but also into which transmission line it is routed. Since the router is operated quantum-mechanically, this allows for transmission of photons in a superposition of paths if the control photon is in a superposition of states. Because of the latter, the proposed router can also be used to construct a quantum random access memory~\cite{Lloyd}, where a tree-like network with quantum switches at the nodes leads an address register of qubits from the root node to a superposition of memory cells. Finally, the router could be used as a basic element in quantum communication networks. A list of requirements to fulfill this purpose is discussed in the Supplemental Material~\cite{Note1}.

This work is part of the research programme of the Foundation for Fundamental Research on Matter (FOM), which is part of the Netherlands Organisation for Scientific Research (NWO).

%


\pagebreak
\widetext
\begin{center}
\textbf{\large Proposal for a transmon-based quantum router: Supplemental Material}
\end{center}

\renewcommand{\theequation}{S\arabic{equation}}
\renewcommand{\thefigure}{S\arabic{figure}}
\renewcommand{\thetable}{S\arabic{table}}

\vspace{1cm}
These notes contain some comments and additional information needed to understand the derivation of the results presented in the main text. We discuss in detail the need for a  nonlinear capacitor in the proposed router design and possible candidates together with their properties. We also show how our model (Hamiltonian) has been derived and quantized and how scattering amplitudes and probabilities are computed. We present the set of numbers we have used and discuss obstacles and possible remedies towards scaling up of our proposed router. Finally, we discuss the requirements that we believe a quantum router needs to satisfy in order to construct a network for quantum communications.

\section{Nonlinear capacitors}

In circuit QED, a device similar to the one proposed in the main text (see also FIG.~\ref{fig:deviceS}) is usually described by the Jaynes-Cummings (or Tavis-Cummings) Hamiltonian~\cite{Sdicarlo,Slucero,Skoch}. This Hamiltonian only allows the excitation of one level of the transmon by absorbing one photon. If the third level of the transmon is to be reached, then three photons are needed. Here we propose a mechanism with which the third level of the transmon can be excited in one step by absorbing one photon. Unfortunately this transmon-cavity interaction cannot be achieved by using Josephson junctions or any other inductive element (when including an inductive element in the Lagrangian and performing a Legendre transformation, the resulting Hamiltonian does not contain a term coupling the transmon with the transmission lines, as is required). The nonlinearity provided by a Josephson junction is insufficient to couple the third level of the transmon to the transmission lines in a way that allows for a photon to be absorbed by the system and drive the transmon from its ground state to its third excited state ($\left| T3 \right\rangle$ in the main text). This photon-transmon interaction, not usually present in cQED devices, can, however, be achieved by using a nonlinear capacitative element. These capacitors have been realized and studied in multiple disciplines and with different implementations, such as using ferroelectric thin films or ceramics~\cite{SAraujo,SGluskin}, quantum wells in heterojunctions~\cite{Ssengouga}, MOS junctions~\cite{SColinge,SKhan} and carbon nanotubes~\cite{SIlani,SAkinwande2}.

We are interested in a nonlinear capacitor whose energy is a non-quadratic, symmetric and increasing function of the voltage for small fluctuations of $V$. In this way, the system will be confined into a region of small potentials~\cite{NoteS1}.

Among the aforementioned nonlinear capacitors, the one based on carbon nanotubes seems to be the most suitable candidate. This capacitor~\cite{SIlani} consists of two parallel plates with carbon nanotubes in between, placed perpendicular to the plates. The nonlinearity in the $C(V)$ curve is due to the finiteness of the density of states (DOS) in the nanotubes~\cite{SIlani,SAkinwande2}. The capacitance of this device goes as $C(V)\sim C_0(1 + \alpha V^2)$, thus the energy spectrum of this capacitor as a function of the potential difference across the plates has the form
\begin{align} \label{eq:nlc}
E(V) \sim \int_0^V dV'(V-V') C(V') = \frac{1}{2} C_0 (V^2 + \alpha V^4),
\end{align}
where, to simplify the notation, the constant $\alpha$ has been redefined to absorb a numerical constant.

Note that the carbon nanotubes do not constitute a key element in our proposal. Any other capacitative element, easier to fabricate and implement, with the same or similar energy spectrum will work as well.

\section{The model: derivation}

\begin{figure}
\includegraphics[width=0.7\linewidth]{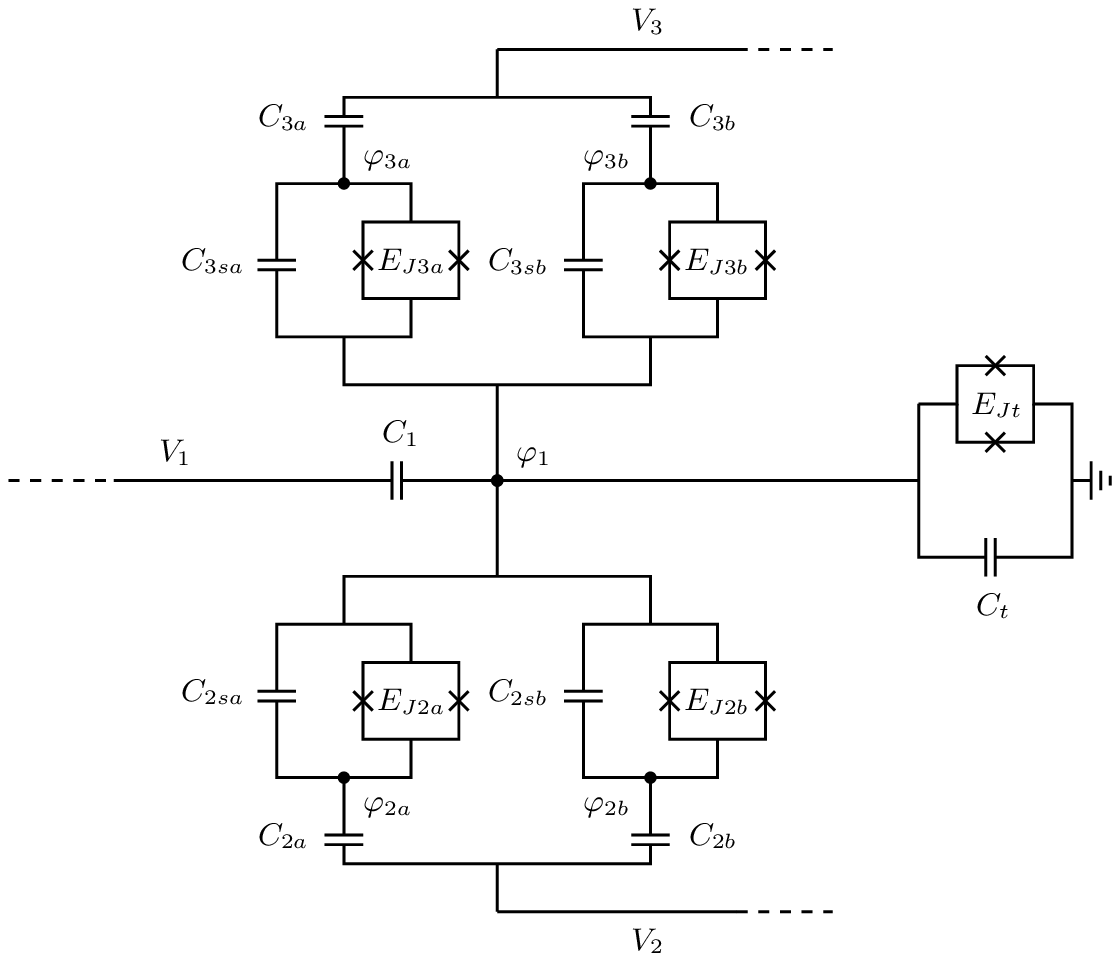}
\caption{Circuit QED device proposed to operate as a quantum router. This device is composed of four SQUIDs, each with a capacitor ($C_{2sa},~C_{2sb},~C_{3sa},~C_{3sb}$) and a pair of Josephson junctions (with energies $E_{J2a}$, etc.). It also contains a transmon qubit, with a capacitor ($C_t$) and a pair of Josephson junctions ($E_{Jt}$). These five elements are capacitively coupled to an incoming transmission line ($V_1$) and to two outgoing transmission lines ($V_2$ and $V_3$).  The fluxes $\varphi_i$, in the nodes of the circuit, are quantized variables (see the text around the figure, section {\em The model: derivation}, for the quantization of these variables) and describe the absorption and emission of incoming photons by the transmon and SQUIDs. In this proposal, two SQUIDs are placed in each branch for scalability purposes and robustness. Since the target photons can be in two states with different frequencies (for scaling purposes, for these will be the control photons in a subsequent router), two resonators (per branch) coupled to the same transmission line, with two distinct threshold energies for which they are detuned, are needed to absorb the target photons. \label{fig:deviceS}}
\end{figure}

The Lagrangian describing our circuit QED system is found in the same way as in circuit theory~\cite{Sbishop,Shouches}. The Lagrangian of the device presented in FIG.~\ref{fig:deviceS} is given by $L = L_S + L_t$, where $L_S$ is the Lagrangian of the SQUIDs, given by
\begin{align}\label{eq:ls}
L_S =&      \frac{C_{2a}}{2} \left( \dot{\varphi}_{2a} - V_2 \right)^2 + \frac{C_{2sa}}{2} \left( \dot{\varphi}_{1} - \dot{\varphi}_{2a} \right)^2 + E_{J2a} \cos\left( \frac{\varphi_1 - \varphi_{2a}}{\varphi_0} \right) \notag \\
&           + \frac{C_{2b}}{2} \left( \dot{\varphi}_{2b} - V_2 \right)^2 + \frac{C_{2sb}}{2} \left( \dot{\varphi}_{1} - \dot{\varphi}_{2b} \right)^2 + E_{J2b} \cos\left( \frac{\varphi_1 - \varphi_{2b}}{\varphi_0} \right) \notag \\
&           + \frac{C_{3a}}{2} \left( \dot{\varphi}_{3a} - V_3 \right)^2 + \frac{C_{3sa}}{2} \left( \dot{\varphi}_{1} - \dot{\varphi}_{3a} \right)^2 + E_{J3a} \cos\left( \frac{\varphi_1 - \varphi_{3a}}{\varphi_0} \right) \notag \\
&           + \frac{C_{3b}}{2} \left( \dot{\varphi}_{3b} - V_3 \right)^2 + \frac{C_{3sb}}{2} \left( \dot{\varphi}_{1} - \dot{\varphi}_{3b} \right)^2 + E_{J3b} \cos\left( \frac{\varphi_1 - \varphi_{3b}}{\varphi_0} \right),
\end{align}
where the dynamic variables $\varphi_1$, $\varphi_{2a}$, $\varphi_{2b}$, $\varphi_{3a}$ and $\varphi_{3b}$ are the fluxes defined at each node of the diagram in FIG.~\ref{fig:deviceS} and $\varphi_0=\hbar/(2e)$ is the flux quantum divided by $2\pi$. For the transmon, the Lagrangian that describes its behavior is given by
\begin{align}\label{eq:lt}
L_t =&      \frac{C_1}{2} \left( \dot{\varphi}_t - V_1 \right)^2 + \frac{C_t}{2} \left( \dot{\varphi}_t^2 + \alpha \dot{\varphi}_1^4 \right) + E_{Jt} \cos\left( \frac{\varphi_1}{\varphi_0} \right).
\end{align}
Note that the expression in Eq.~\eqref{eq:lt} contains the energy spectrum of a carbon nanotube nonlinear capacitor [right-hand side in Eq.~\eqref{eq:nlc}].

The Legendre transformation of the Lagrangian in Eq.~\eqref{eq:ls} and \eqref{eq:lt}, needed to study the quantum dynamics of the router, gives a function that is not easy to work with due to the presence of quadratic and cubic radicals. Instead, a Taylor expansion of the transformation is more suitable for our purposes. The expanded Hamiltonian is given by:
\begin{align}\label{eq:htaylor}
H =&        \frac{p^2_{2a} + 2p_{2a} c_{2a} V_2 - C_{2sa} C_{2a} V^2_2}{2(C_{2a} + C_{2sa})} + \frac{p^2_{2b} + 2p_{2b} c_{2b} V_2 - C_{2sb} C_{2b} V^2_2}{2(C_{2b} + C_{2sb})} \notag \\
&           + \frac{p^2_{3a} + 2p_{3a} c_{3a} V_3 - C_{3sa} C_{3a} V^2_3}{2(C_{3a} + C_{3sa})} + \frac{p^2_{3b} + 2p_{3b} c_{3b} V_3 - C_{3sb} C_{3b} V^2_3}{2(C_{3b} + C_{3sb})} \notag \\
&           + \frac{x^2}{2\gamma} - \frac{\beta x^4}{12 \gamma^4} + \frac{\beta^2 x^6}{18 \gamma^7} - \frac{\beta^3 x^8}{18 \gamma^{10}} - \frac{C_1 V_1^2}{2} - E_{Jt} \cos\left( \frac{\varphi_1}{\varphi_0} \right) \notag \\
&           - E_{J2a} \cos\left( \frac{\varphi_1 - \varphi_{2a}}{\varphi_0} \right) - E_{J2b} \cos\left( \frac{\varphi_1 - \varphi_{2b}}{\varphi_0} \right) \notag \\
&           - E_{J3a} \cos\left( \frac{\varphi_1 - \varphi_{3a}}{\varphi_0} \right) - E_{J3b} \cos\left( \frac{\varphi_1 - \varphi_{3b}}{\varphi_0} \right).
\end{align}
For the sake of simplicity, the Hamiltonian in Eq.~\eqref{eq:htaylor} is expressed in terms of the fluxes, their conjugate momenta and the variable $x$, defined as
\begin{align}\label{eq:x}
x \equiv &~   p_1 + C_1 V_1 + \frac{C_{2sa}}{C_{2a} + C_{2sa}} (p_{2a} + C_{2a} V_2) + \frac{C_{2sb}}{C_{2b} + C_{2sb}} (p_{2b} + C_{2b} V_2) \notag \\
&     + \frac{C_{3sa}}{C_{3a} + C_{3sa}} (p_{3a} + C_{3a} V_3) + \frac{C_{3sb}}{C_{3b} + C_{3sb}} (p_{3b} + C_{3b} V_3),
\end{align}
with
\begin{align}
\beta \equiv &~    6\, C_t \alpha \\
\gamma \equiv &~   C_1 + C_t + \frac{C_{2a} C_{2sa}}{C_{2a} + C_{2sa}} + \frac{C_{2b} C_{2sb}}{C_{2b} + C_{2sb}} + \frac{C_{3a} C_{3sa}}{C_{3a} + C_{3sa}} + \frac{C_{3b} C_{3sb}}{C_{3b} + C_{3sb}}.
\end{align}
In order to give rise to a density-density interaction between the different energy levels, the Hamiltonian is required to contain a term with $p_1^6 p_{2a}^2$ (see the next section in which the Hamiltonian is quantized). This expression appears after expanding the Hamiltonian in a Taylor series. In this same expansion a term containing $p_1^3 V_1$ appears which describes the coupling between the third energy level of the transmon and the transmission lines.

After substituting Eq.~\eqref{eq:x} in Eq.~\eqref{eq:htaylor}, many more interactions appear, but they can be neglected by imposing constraints on the parameters of the system. To start with, the coupling between the transmon and the outgoing transmission lines can be neglected by making $C_{2sa}$, $C_{2sb}$, $C_{3sa}$ and $C_{3sb}$ smaller than $C_1$, $C_{2a}$, $C_{2b}$, $C_{3a}$ and $C_{3b}$. In this way, the exchange interaction between the SQUID $2a$ and the third outgoing transmission line is also negligible. In order to see how the remaining couplings are enhanced or neglected we first have to quantize the Hamiltonian.

\subsection{Quantization of the Hamiltonian}

The Hamiltonian is quantized as usual by introducing the ladder operators as follows:
\begin{align}
p_1 =& -\frac{i\hbar}{2\varphi_0} \left( \frac{3}{2} \frac{\gamma^2}{\beta E_T} \right)^{1/4} (a_1 - a^\dagger_1) \label{eq:qp1} \\
\phi_1 =& \left( \frac{2}{3} \frac{\beta E_T}{\gamma^2} \right)^{1/4} (a_1 + a^\dagger_1) \label{eq:qph1}.
\end{align}
Here we have made use of the variable $\phi_1=\varphi_1/\varphi_0$, to simplify the derivations, where $\varphi_0=\frac{\hbar}{2e}$ and also $E_T=\frac{\hbar^2}{8\gamma \varphi^2_0}$. The operators $a_1$ and $a^\dagger_1$ are the usual ladder operators that annihilate and create an excitation in the transmon. For the SQUIDs, the quantized forms of the operators are
\begin{align}
p_{2a} =& -\frac{i\hbar}{2\varphi_0} \left[ E_{J2a} \frac{C_{2a} + C_{2sa}}{C_{2sa}} \left( \frac{\beta}{6 \gamma^2 E_T} \right)^{1/2} \right]^{1/2} (a_{2a} - a^\dagger_{2a}) \label{eq:qp2a} \\
\phi_{2a} =& \left[ \frac{1}{E_{J2a}} \frac{C_{2sa}}{C_{2a} + C_{2sa}} \left( \frac{6 \gamma^2 E_T}{\beta} \right)^{1/2} \right]^{1/2} (a_1 + a^\dagger_1) \label{eq:qph2a}.
\end{align}
Similar expressions can be derived for the other operators. With Eqns.~\eqref{eq:qp1} -- \eqref{eq:qph2a} together with $E_{Jt} + E_{J2a} + E_{J2b} + E_{J3a} + E_{J3b} = 3\gamma^2 /\beta \equiv \bar{E}_J$ and $E_{J2a} E_{2a} = (\bar{E}_J - E_{J2a} + E_{J2b} + E_{J3a} + E_{J3b})$ (for simplicity we have also introduced $E_{2a}=\frac{\hbar^2}{8C_{2sa} \varphi^2_0}$) the SQUID-transmon exchange interactions are strongly suppressed and the expressions describing the energy levels of the system together with the density-density interactions between them are found [see Eq.~(2) in the main text]. The SQUID-SQUID exchange interaction can be neglected since its interaction strength goes as $(C_{2sa}/C_{2a})^2$. Other similar interactions can be neglected by making the capacitances $C_{2sa}$, $C_{2sb}$, $C_{3sa}$ and $C_{3sb}$ smaller than all the others and using the rotating wave approximation.

The interaction Hamiltonian [Eq.~(3) in the main text], which couples the transmon and the SQUIDs with the incoming and outgoing transmission lines, is derived by imposing the following quantization of the potentials $V_1$, $V_2$ and $V_3$:
\begin{align}\label{eq:qvi}
V_i = -\frac{i\hbar}{2 \varphi_0} A_i^{1/4} (b_i - b^\dagger_i).
\end{align}
Here $b_i,~ b^\dagger_i$ are the ladder operators describing the oscillating modes in the $i$-th transmission line, and $A_i$ are three constants that may be different.

Let us look again the expressions in Eqns.~\eqref{eq:htaylor} and \eqref{eq:x}. In the Hamiltonian [Eq.~\eqref{eq:htaylor}], there are only even powers of $x$. If we consider a single incoming photon, contained in $V_1$ [see Eq.~\eqref{eq:qvi}], given that the terms in the Hamiltonian containing a single power of $V_1$ only contains odd powers of the momenta, the photon will excite only odd levels of the system. An even level can only be excited if an odd level is already occupied. This is shown in the Hamiltonian $H_c$ in the main text [Eq.~(3)].


\subsection{Projection operators}

In order to simpliby both the notation and the calculations we have introduced three projection operators defined as
\begin{align}
a^\dagger_{T1} \equiv & \frac{1}{6} a_1^2 {a^\dagger_1}^3 = \left( \begin{array}{cccc} 0&0&0&0\\0&0&0&0\\0&0&0&1\\0&0&0&0 \end{array} \right), \label{eq:defa1} \\
a^\dagger_{T2} \equiv & \frac{1}{3\sqrt{2}} a_1 {a^\dagger_1}^3 = \left( \begin{array}{cccc} 0&0&0&0\\0&0&0&1\\0&0&0&0\\0&0&0&0 \end{array} \right), \label{eq:defa2} \\
a^\dagger_{T3} \equiv & \frac{1}{\sqrt{6}} {a^\dagger_1}^3 = \left( \begin{array}{cccc} 0&0&0&1\\0&0&0&0\\0&0&0&0\\0&0&0&0 \end{array} \right) \label{eq:defa3},
\end{align}
to replace $a_1$ and its hermitian conjugate. The Hamiltonian in the main text is written in terms of these new operators (see the section {\em Analysis}). These operators have the property that $a^\dagger_{Ti}$ acting on the ground state of the transmon creates an excitation in the $i$-th level, whereas the same operator acting on any other state where the transmon is already excited gives zero. Similarly, the operator $a_{Ti}$ only annihilates an excitation of levels in the transmon, lowering it to its ground state.

With these projection operators, the quantized Hamiltonian $H_c$, which describes the coupling between the transmission lines with the transmon and SQUIDs takes the simple form
\begin{align}
H_c=& \int dp \left[ \frac{a^\dagger_{T1} b_1(p)}{\sqrt{\pi \tau_{T1}}}
 + \frac{a^\dagger_{T3} b_1(p)}{\sqrt{\pi \tau_{T3}}}
 + \left( \sqrt{\frac{2}{\pi \tau_{T1}}} - \sqrt{\frac{3}{\pi \tau_{T3}}} \right) a^\dagger_{T2} a_{T1} b_1(p)
 + \left( \sqrt{\frac{3}{\pi \tau_{T1}}} - 3\sqrt{\frac{2}{\pi \tau_{T3}}} \right) a^\dagger_{T3} a_{T2} b_1(p) \right. \notag \\
&     \qquad + \left. \left( \frac{a^\dagger_{2a}}{\sqrt{\pi \tau_a}} + \frac{a^\dagger_{2b}}{\sqrt{\pi \tau_b}} \right) \left( b_1(p) + b_2(p) \right)) + 
\left( \frac{a^\dagger_{3a}}{\sqrt{\pi \tau_a}} + \frac{a^\dagger_{3b}}{\sqrt{\pi \tau_b}} \right) \left( b_1(p) + b_3(p) \right)) + h.c. \right].
\end{align}
The fourth term in the right-hand side, whose hermitian conjugate describes the process where the third excited level of the transmon decays into the second while emitting a photon, can be canceled by setting $\alpha$ as
\begin{align}\label{eq:alpha}
\alpha = \frac{\gamma^3 \varphi_0^2}{324 \hbar C_t}.
\end{align}
In this way, the number of states the third level of the transmon can evolve into is reduced and, thus, the lifetime of the third level of the transmon is enhanced, increasing the transition matrix element from the ground state to $\left| T3 \right\rangle$.

\subsection{Langevin equations of motion}

The Langevin equations of the `static' variables of the system in terms of $b_{1in}$, $b_{2in}$ and $b_{3in}$ are
\begin{align}
\dot{a}_{T1} =&   -\left( i\omega_{T1} + \frac{1}{\tau_{T1}} \right) a_{T1}
 - \frac{1}{\sqrt{\tau_{T1} \tau_{T3}}} a_{T3}
 + i\sum_k J_{1k} a_{T1} a^\dagger_k a_k
 + 2 \left( \sqrt{\frac{2}{\tau_{T1}}} - \sqrt{\frac{3}{\tau_{T3}}} \right) \left( \frac{a^\dagger_{T1} a_{T2}}{\sqrt{\tau_{T1}}} + \frac{a^\dagger_{T3} a_{T2}}{\sqrt{\tau_{T3}}} \right) \notag \\
 &                + \left[ \frac{1}{\sqrt{\tau_{T1}}} \left( a^\dagger_{T1} a_{T1} - a_{T1}a^\dagger_{T1} \right) + \frac{1}{\sqrt{\tau_{T3}}} a^\dagger_{T3} a_{T1} \right] \left( \frac{a_{2a}}{\sqrt{\tau_a}} + \frac{a_{2b}}{\sqrt{\tau_b}} + \frac{a_{3a}}{\sqrt{\tau_a}} + \frac{a_{3b}}{\sqrt{\tau_b}} \right) \notag \\
&                 + \left( \sqrt{\frac{2}{\tau_{T1}}} - \sqrt{\frac{3}{\tau_{T3}}} \right) \left( \frac{a^\dagger_{2a}}{\sqrt{\tau_a}} + \frac{a^\dagger_{2b}}{\sqrt{\tau_b}} + \frac{a^\dagger_{3a}}{\sqrt{\tau_a}} + \frac{a^\dagger_{3b}}{\sqrt{\tau_b}} \right) a_{T2} \notag \\
&                 + i \left[ \sqrt{\frac{2}{\tau_{T1}}} \left( a^\dagger_{T1} a_{T1} - a_{T1} a^\dagger_{T1} \right) + \sqrt{\frac{2}{\tau_{T3}}} a^\dagger_{T3} a_{T1} \right] b_{1in}(t) - i\sqrt{2} \left( \sqrt{\frac{2}{\tau_{T1}}} - \sqrt{\frac{3}{\tau_{T3}}} \right) b^\dagger_{1in}(t) a_{T2},\label{eq:at1} \\
\dot{a}_{T3} =&   -\left( i\omega_{T3} + \frac{1}{\tau_{T3}} \right) a_{T3}
 - \frac{1}{\sqrt{\tau_{T1} \tau_{T3}}} a_{T1}
 + i\sum_k J_{3k} a_{T3} a^\dagger_k a_k \notag \\
 &                + \left[ \frac{1}{\sqrt{\tau_{T1}}} a^\dagger_{T1} a_{T3} + \frac{1}{\sqrt{\tau_{T3}}} \left( a^\dagger_{T3} a_{T3} - a_{T3}a^\dagger_{T3} \right) \right] \left( \frac{a_{2a}}{\sqrt{\tau_a}} + \frac{a_{2b}}{\sqrt{\tau_b}} + \frac{a_{3a}}{\sqrt{\tau_a}} + \frac{a_{3b}}{\sqrt{\tau_b}} \right) \notag \\
&                 + i \left[ \sqrt{\frac{2}{\tau_{T1}}} a^\dagger_{T1} a_{T3} + \sqrt{\frac{2}{\tau_{T3}}} \left( a^\dagger_{T3} a_{T3} - a_{T3} a^\dagger_{T3} \right) \right] b_{1in}(t), \label{eq:at3} \\
\dot{a}_{2a} =&    -\left( i\omega_{2a} + \frac{2}{\tau_a} \right) a_{2a}
 + i\sum_{j=1}^3 J_{j2a}a^\dagger_{Tj} a_{Tj} a_{2a}
 - i\sqrt{\frac{2}{\tau_a}} \left( b_{1in}(t) + b_{2in}(t) \right) \notag \\
&                 - \frac{2 a_{2b}}{\sqrt{\tau_a \tau_b}} - \frac{a_{3a}}{\tau_a} - \frac{a_{3b}}{\sqrt{\tau_a \tau_b}}
 - \frac{1}{\sqrt{\tau_a}} \left( \sqrt{\frac{2}{\tau_{T1}}} - \sqrt{\frac{3}{\tau_{T3}}} \right)a^\dagger_{T1} a_{T2} - \frac{1}{\sqrt{\tau_a \tau_{T1}}} a_{T1} - \frac{1}{\sqrt{\tau_a \tau_{T3}}} a_{T3}, \label{eq:a2a}
\end{align}
where the index $k$ in the sum runs over $k \in \{ 2a,2b,3a,3b \}$ and three more equations for $2b$, $3a$ and $3b$, similar to Eq.~\eqref{eq:a2a}, have to be derived. Relaxation and dephasing ratios are then introduced into this set of Langevin equations by following Refs.~\cite{Sgardiner,Sithier}.


\section{Photon modes in the transmission lines}

In this section we briefly describe the meaning of the left/right-moving and even/odd modes, which are common in many circuit-QED works where the input-output formalism is used. For more details we refer the reader to Ref.~\cite{Sfan}, appendix A, where the following description is taken from.

In case the photons in a transmission line are allowed to propagate in both directions, the photon modes are described by right-moving bosonic operators ($r$, $r^\dagger$) if they propagate to the right or by left-moving bosonic operators ($l$, $l^\dagger$) if they propagate to the left. In this case, the Hamiltonian that describes the transmission lines reads:
\begin{align}
H_T = \int_{-\infty}^{\infty} dp\, p \left( r^\dagger_p r_p - l^\dagger_p l_p \right),
\end{align}
for each transmission line. The minus sign appears because the dispersion relation of the left-moving modes is the mirror image of the one of the right-moving modes. The remainder of the Hamiltonian given in the main text should, in this case, contain the sum of operators $r+l$ and $r^\dagger + l^\dagger$ instead of $b$ and $b^\dagger$ for each transmission line, so the different components of the router interact equally with both the left and right-moving modes.

From this new Hamiltonian, and following the same procedure as before (input/output formalism), similar equations of motion are found. Nevertheless, we can also do one more step and define two new operators: the even ($b$, $b^\dagger$) and odd ($\mathring{b}$, $\mathring{b}^\dagger$) operators that describe what are called even and odd modes, defined as:
\begin{align}
b_p =& \frac{r_p + l_{-p}}{\sqrt{2}},\\
\mathring{b}_p =& \frac{r_p - l_{-p}}{\sqrt{2}}.
\end{align}
These we then introduce in the Hamiltonian replacing the right and left-moving modes. One can check that the odd modes are not coupled to the system (they do not appear in the interacting part of the Hamiltonian), so they can be omitted in the Hamiltonian. The even modes are the ones used in the main text.

\section{Numerical integration of the scattering amplitudes}

In order to find the set of values for the capacitances ---these are the only free parameters left--- that best satisfy the requirements for an optimal quantum router it is convenient to maximize (or minimize) the scattering probabilities according to the operating requirements. Since an analytical solution has proved difficult to obtain, we have looked, by trial and error, for a set of numbers with which the router works as expected and found a set of capacitances that satisfy our requirements of energy spacing between the SQUIDs and transmon levels. Then, by maximizing the scattering amplitudes we have fine-tuned these values. Although a more systematic way of searching may result in a better solution, we have found that there exist at least one set of such parameters, shown in Table~\ref{tab:cap}. The reason for this fine tuning is that, although the separation between the energy levels of the transmon is relatively large, the separation between the energy levels of the SQUIDs is very small (and the optimal operation of the quantum router sensitively depends on small variations of the values of the capacitances)

\begin{table}
\begin{ruledtabular}
\begin{tabular}{c c c c c c c c c}
$C_t$ & $C_{2a}$ & $C_{2b}$ & $C_{3a}$ & $C_{3b}$ & $C_{2sa}$ & $C_{2sb}$ & $C_{3sa}$ & $C_{3sb}$ \\ \hline
0.1 $C_1$ & 1.2 $C_1$ & 1.1 $C_1$ & $C_1$ & $C_1$ & 0.1824 $C_1$ & 0.1934 $C_1$ & 0.1560 $C_1$ & 0.1799 $C_1$
\end{tabular}
\end{ruledtabular}
\caption{Proposal for a set of parameters that satisfy the requirements of a quantum router, given in terms of the capacitance $C_1=10^{-9} \frac{e^2}{\hbar} F$. With these values, the device can operate as described in the main text, forwarding photons into one of the outgoing transmission lines according to the state of the control photon. The rest of the parameters (Josephson energies, inductances, $\alpha$) are found as functions of the capacitances during the derivation of the quantized Hamiltonian (see the second section of these supplementary notes)~\label{tab:cap}.}
\end{table}

With the values in Table~\ref{tab:cap}, the energies are found to be around $10\,GHz$ for the transmon energy levels ($\omega_{T1}=7.189\, GHz$ and $\omega_{T3}=21.93\, GHz$ ) and around $1\, GHz$ for the SQUIDs ($\omega_{2a}=1.028\, GHz$, $\omega_{2b}=1.162\, GHz$, $\omega_{3a}=1.053\, GHz$ and $\omega_{3b}=1.186\, GHz$). The coupling strength is of the order of $10^{-7}\, Hz$. All these values are common in the related literature. Also, we used the ratio $\bar{E}_J/E_T=1296$ which, although being larger than in other circuit-QED works, is sufficient to give rise to the nonlinear effects needed for the correct operation of the quantum router~\cite{Skoch}.

For the two-photon processes, we have made some assumptions: from the single-photon process, we have seen that the transmon absorbs the first incoming photon with a large probability, so for the two-photon processes we assumed that we sent the second photon while the transmon is already excited and that it does not decay emitting the control photon until the second photon has been fully transmitted. This implies that the lifetime of the transmon is much larger than that of the SQUIDs. This is a plausible assumption, given that the lifetimes of such cQED devices depend not only on the elements they are composed of but also on the geometry of the system (Purcell effect, see e.g.~\cite{Sreed}).

Finally, we comment on the possibility to use the quantum router in a larger network containing many quantum routers such that the transmitted photons of one router can be used as control photons in the next node of the network. This means that a photon transmitted by the router, which has enough energy to excite one of the SQUIDs, can also excite the transmon in the next router. Therefore, in a scalable network where all the routers are identical, the energy needed to excite the transmon ($\omega_{T1}$ or $\omega_{T3}$) is the same as the energy needed to excite one of the SQUIDs ($\omega_a$ or $\omega_b$). In our (optimized) model, however, the energy of the first excited level of the transmon is one order of magnitude larger than that of the SQUIDs with the current set of capacitances. A possible solution may be to add a device between the two consecutive routers that enlarges the energy of the first photon to arrive. Stimulated Raman Adiabatic Passage (STIRAP) processes may be a candidate for a device capable of transferring the population of one quantum state (with lower energy) to another (with higher energy). These processes have been studied in circuit QED systems in~\cite{SSiewert,SFeng}.

\section{Requirements for an efficient quantum router for quantum communications}

So far, different proposals for quantum switches or routers that may be implemented in a quantum network have been studied~\cite{Sagarwal,Shoi,Slu,Slemr,Sgarcia}. In order to assess the suitability of a given implementation for efficient use in a quantum communication network, we list here a set of requirements that we believe a quantum router must satisfy (part of these have been proposed elsewhere~\cite{Slemr,Sgarcia}). The list below contains general requirements, which can be applied and checked for specific implementations.

Imagine we have a network for quantum communication composed of quantum channels (which allow for the transmission of quantum information) and nodes (which contain quantum routers). Consider also that there is a signal to be sent and an element that controls the switches.
\begin{enumerate}
\item The information in the signal to be transmitted and also the information encoded in the control elements must be stored in quantum objects (qubits). Otherwise we can not talk about a fully quantum-mechanical router~\cite{Slemr}.
\item The router has to be able to route the (quantum) signal into a coherent superposition of both output modes~\cite{Slemr}.
\item The signal information must remain undisturbed during the entire routing process~\cite{Slemr}. The control information must also remain undisturbed, so that we can keep track of the signal and verify that it is not modified (due to e.g., entanglement with the control element).
\item The router has to work without any need for postselection on the signal output~\cite{Slemr}.
\item To optimize the resources of the quantum network, only a single control qubit is required to direct each signal qubit~\cite{Slemr}. Alternatively, the control information could also be stored in the signal qubit itself~\cite{Sgarcia}, but that may give the receiver information about the network which, in some applications, may not be a good idea.
\item With one single control qubit, as many as possible signal qubits should be forwarded.
\item A router with $n$ outputs must operate with the same efficiency (or speed, or with the same number of operations, etc.) for each output channel.
\item Scalability: the output of a router should be the input of the next one in a network.
\end{enumerate}

The quantum router that we propose satisfies all of these requirements except for the last one (although, as we discussed in the previous section, this problem can for example be amended by using STIRAP processes). We thus believe that it can become a good candidate for a quantum switch or router in a network for quantum communication.


\begin{thebibliography}{42}%
\makeatletter
\providecommand \@ifxundefined [1]{%
 \@ifx{#1\undefined}
}%
\providecommand \@ifnum [1]{%
 \ifnum #1\expandafter \@firstoftwo
 \else \expandafter \@secondoftwo
 \fi
}%
\providecommand \@ifx [1]{%
 \ifx #1\expandafter \@firstoftwo
 \else \expandafter \@secondoftwo
 \fi
}%
\providecommand \natexlab [1]{#1}%
\providecommand \enquote  [1]{``#1''}%
\providecommand \bibnamefont  [1]{#1}%
\providecommand \bibfnamefont [1]{#1}%
\providecommand \citenamefont [1]{#1}%
\providecommand \href@noop [0]{\@secondoftwo}%
\providecommand \href [0]{\begingroup \@sanitize@url \@href}%
\providecommand \@href[1]{\@@startlink{#1}\@@href}%
\providecommand \@@href[1]{\endgroup#1\@@endlink}%
\providecommand \@sanitize@url [0]{\catcode `\\12\catcode `\$12\catcode
  `\&12\catcode `\#12\catcode `\^12\catcode `\_12\catcode `\%12\relax}%
\providecommand \@@startlink[1]{}%
\providecommand \@@endlink[0]{}%
\providecommand \url  [0]{\begingroup\@sanitize@url \@url }%
\providecommand \@url [1]{\endgroup\@href {#1}{\urlprefix }}%
\providecommand \urlprefix  [0]{URL }%
\providecommand \Eprint [0]{\href }%
\providecommand \doibase [0]{http://dx.doi.org/}%
\providecommand \selectlanguage [0]{\@gobble}%
\providecommand \bibinfo  [0]{\@secondoftwo}%
\providecommand \bibfield  [0]{\@secondoftwo}%
\providecommand \translation [1]{[#1]}%
\providecommand \BibitemOpen [0]{}%
\providecommand \bibitemStop [0]{}%
\providecommand \bibitemNoStop [0]{.\EOS\space}%
\providecommand \EOS [0]{\spacefactor3000\relax}%
\providecommand \BibitemShut  [1]{\csname bibitem#1\endcsname}%
\let\auto@bib@innerbib\@empty
\bibitem [{\citenamefont {H\"{a}ffner}\ \emph {et~al.}(2008)\citenamefont
  {H\"{a}ffner}, \citenamefont {Roos},\ and\ \citenamefont {Blatt}}]{haffner}%
  \BibitemOpen
  \bibfield  {author} {\bibinfo {author} {\bibfnamefont {H.}~\bibnamefont
  {H\"{a}ffner}}, \bibinfo {author} {\bibfnamefont {C.}~\bibnamefont {Roos}}, \
  and\ \bibinfo {author} {\bibfnamefont {R.}~\bibnamefont {Blatt}},\ }\href
  {\doibase http://dx.doi.org/10.1016/j.physrep.2008.09.003} {\bibfield
  {journal} {\bibinfo  {journal} {Physics Reports}\ }\textbf {\bibinfo {volume}
  {469}},\ \bibinfo {pages} {155 } (\bibinfo {year} {2008})}\BibitemShut
  {NoStop}%
\bibitem [{\citenamefont {DiCarlo}\ \emph {et~al.}(2009)\citenamefont
  {DiCarlo}, \citenamefont {Chow}, \citenamefont {Gambetta}, \citenamefont
  {Bishop}, \citenamefont {Johnson}, \citenamefont {Schuster}, \citenamefont
  {Majer}, \citenamefont {Blais}, \citenamefont {Frunzio}, \citenamefont
  {Girvin},\ and\ \citenamefont {Schoelkopf}}]{dicarlo}%
  \BibitemOpen
  \bibfield  {author} {\bibinfo {author} {\bibfnamefont {L.}~\bibnamefont
  {DiCarlo}}, \bibinfo {author} {\bibfnamefont {J.~M.}\ \bibnamefont {Chow}},
  \bibinfo {author} {\bibfnamefont {J.~M.}\ \bibnamefont {Gambetta}}, \bibinfo
  {author} {\bibfnamefont {L.~S.}\ \bibnamefont {Bishop}}, \bibinfo {author}
  {\bibfnamefont {B.~R.}\ \bibnamefont {Johnson}}, \bibinfo {author}
  {\bibfnamefont {D.~I.}\ \bibnamefont {Schuster}}, \bibinfo {author}
  {\bibfnamefont {J.}~\bibnamefont {Majer}}, \bibinfo {author} {\bibfnamefont
  {A.}~\bibnamefont {Blais}}, \bibinfo {author} {\bibfnamefont
  {L.}~\bibnamefont {Frunzio}}, \bibinfo {author} {\bibfnamefont {S.~M.}\
  \bibnamefont {Girvin}}, \ and\ \bibinfo {author} {\bibfnamefont {R.~J.}\
  \bibnamefont {Schoelkopf}},\ }\href {\doibase 10.1038/nature08121} {\bibfield
   {journal} {\bibinfo  {journal} {Nature}\ }\textbf {\bibinfo {volume}
  {460}},\ \bibinfo {pages} {240} (\bibinfo {year} {2009})}\BibitemShut
  {NoStop}%
\bibitem [{\citenamefont {van~der Sar}\ \emph {et~al.}(2012)\citenamefont
  {van~der Sar}, \citenamefont {Wang}, \citenamefont {Blok}, \citenamefont
  {Bernien}, \citenamefont {Taminiau}, \citenamefont {Toyli}, \citenamefont
  {Lidar}, \citenamefont {Awschalom}, \citenamefont {Hanson},\ and\
  \citenamefont {Dobrovitski}}]{vandersar}%
  \BibitemOpen
  \bibfield  {author} {\bibinfo {author} {\bibfnamefont {T.}~\bibnamefont
  {van~der Sar}}, \bibinfo {author} {\bibfnamefont {Z.~H.}\ \bibnamefont
  {Wang}}, \bibinfo {author} {\bibfnamefont {M.~S.}\ \bibnamefont {Blok}},
  \bibinfo {author} {\bibfnamefont {H.}~\bibnamefont {Bernien}}, \bibinfo
  {author} {\bibfnamefont {T.~H.}\ \bibnamefont {Taminiau}}, \bibinfo {author}
  {\bibfnamefont {D.~M.}\ \bibnamefont {Toyli}}, \bibinfo {author}
  {\bibfnamefont {D.~A.}\ \bibnamefont {Lidar}}, \bibinfo {author}
  {\bibfnamefont {D.~D.}\ \bibnamefont {Awschalom}}, \bibinfo {author}
  {\bibfnamefont {R.}~\bibnamefont {Hanson}}, \ and\ \bibinfo {author}
  {\bibfnamefont {V.~V.}\ \bibnamefont {Dobrovitski}},\ }\href {\doibase
  10.1038/nature10900} {\bibfield  {journal} {\bibinfo  {journal} {Nature}\
  }\textbf {\bibinfo {volume} {484}},\ \bibinfo {pages} {82} (\bibinfo {year}
  {2012})}\BibitemShut {NoStop}%
\bibitem [{\citenamefont {Lucero}\ \emph {et~al.}(2012)\citenamefont {Lucero},
  \citenamefont {Barends}, \citenamefont {Chen}, \citenamefont {Kelly},
  \citenamefont {Mariantoni}, \citenamefont {Megrant}, \citenamefont
  {O'Malley}, \citenamefont {Sank}, \citenamefont {Vainsencher}, \citenamefont
  {Wenner}, \citenamefont {White}, \citenamefont {Yin}, \citenamefont
  {Cleland},\ and\ \citenamefont {Martinis}}]{lucero}%
  \BibitemOpen
  \bibfield  {author} {\bibinfo {author} {\bibfnamefont {E.}~\bibnamefont
  {Lucero}}, \bibinfo {author} {\bibfnamefont {R.}~\bibnamefont {Barends}},
  \bibinfo {author} {\bibfnamefont {Y.}~\bibnamefont {Chen}}, \bibinfo {author}
  {\bibfnamefont {J.}~\bibnamefont {Kelly}}, \bibinfo {author} {\bibfnamefont
  {M.}~\bibnamefont {Mariantoni}}, \bibinfo {author} {\bibfnamefont
  {A.}~\bibnamefont {Megrant}}, \bibinfo {author} {\bibfnamefont
  {P.}~\bibnamefont {O'Malley}}, \bibinfo {author} {\bibfnamefont
  {D.}~\bibnamefont {Sank}}, \bibinfo {author} {\bibfnamefont {A.}~\bibnamefont
  {Vainsencher}}, \bibinfo {author} {\bibfnamefont {J.}~\bibnamefont {Wenner}},
  \bibinfo {author} {\bibfnamefont {T.}~\bibnamefont {White}}, \bibinfo
  {author} {\bibfnamefont {Y.}~\bibnamefont {Yin}}, \bibinfo {author}
  {\bibfnamefont {A.~N.}\ \bibnamefont {Cleland}}, \ and\ \bibinfo {author}
  {\bibfnamefont {J.~M.}\ \bibnamefont {Martinis}},\ }\href {\doibase
  10.1038/nphys2385} {\bibfield  {journal} {\bibinfo  {journal} {Nat. Phys.}\
  }\textbf {\bibinfo {volume} {8}},\ \bibinfo {pages} {719} (\bibinfo {year}
  {2012})}\BibitemShut {NoStop}%
\bibitem [{\citenamefont {Monroe}\ and\ \citenamefont {Kim}(2013)}]{monroe}%
  \BibitemOpen
  \bibfield  {author} {\bibinfo {author} {\bibfnamefont {C.}~\bibnamefont
  {Monroe}}\ and\ \bibinfo {author} {\bibfnamefont {J.}~\bibnamefont {Kim}},\
  }\href {\doibase 10.1126/science.1231298} {\bibfield  {journal} {\bibinfo
  {journal} {Science}\ }\textbf {\bibinfo {volume} {339}},\ \bibinfo {pages}
  {1164} (\bibinfo {year} {2013})}\BibitemShut {NoStop}%
\bibitem [{\citenamefont {Barz}\ \emph {et~al.}(2014)\citenamefont {Barz},
  \citenamefont {Kassal}, \citenamefont {Ringbauer}, \citenamefont {Lipp},
  \citenamefont {Daki\'{c}}, \citenamefont {Aspuru-Guzik},\ and\ \citenamefont
  {Walther}}]{barz}%
  \BibitemOpen
  \bibfield  {author} {\bibinfo {author} {\bibfnamefont {S.}~\bibnamefont
  {Barz}}, \bibinfo {author} {\bibfnamefont {I.}~\bibnamefont {Kassal}},
  \bibinfo {author} {\bibfnamefont {M.}~\bibnamefont {Ringbauer}}, \bibinfo
  {author} {\bibfnamefont {Y.~O.}\ \bibnamefont {Lipp}}, \bibinfo {author}
  {\bibfnamefont {B.}~\bibnamefont {Daki\'{c}}}, \bibinfo {author}
  {\bibfnamefont {A.}~\bibnamefont {Aspuru-Guzik}}, \ and\ \bibinfo {author}
  {\bibfnamefont {P.}~\bibnamefont {Walther}},\ }\href {\doibase
  http://dx.doi.org/10.1038/srep06115} {\bibfield  {journal} {\bibinfo
  {journal} {Sci. Rep.}\ }\textbf {\bibinfo {volume} {4}},\ \bibinfo {pages}
  {6115} (\bibinfo {year} {2014})}\BibitemShut {NoStop}%
\bibitem [{\citenamefont {Stucki}\ \emph {et~al.}(2011)\citenamefont {Stucki},
  \citenamefont {Legré}, \citenamefont {Buntschu}, \citenamefont {Clausen},
  \citenamefont {Felber}, \citenamefont {Gisin}, \citenamefont {Henzen},
  \citenamefont {Junod}, \citenamefont {Litzistorf}, \citenamefont {Monbaron},
  \citenamefont {Monat}, \citenamefont {Page}, \citenamefont {Perroud},
  \citenamefont {Ribordy}, \citenamefont {Rochas}, \citenamefont {Robyr},
  \citenamefont {Tavares}, \citenamefont {Thew}, \citenamefont {Trinkler},
  \citenamefont {Ventura}, \citenamefont {Voirol}, \citenamefont {Walenta},\
  and\ \citenamefont {Zbinden}}]{stucki}%
  \BibitemOpen
  \bibfield  {author} {\bibinfo {author} {\bibfnamefont {D.}~\bibnamefont
  {Stucki}}, \bibinfo {author} {\bibfnamefont {M.}~\bibnamefont {Legré}},
  \bibinfo {author} {\bibfnamefont {F.}~\bibnamefont {Buntschu}}, \bibinfo
  {author} {\bibfnamefont {B.}~\bibnamefont {Clausen}}, \bibinfo {author}
  {\bibfnamefont {N.}~\bibnamefont {Felber}}, \bibinfo {author} {\bibfnamefont
  {N.}~\bibnamefont {Gisin}}, \bibinfo {author} {\bibfnamefont
  {L.}~\bibnamefont {Henzen}}, \bibinfo {author} {\bibfnamefont
  {P.}~\bibnamefont {Junod}}, \bibinfo {author} {\bibfnamefont
  {G.}~\bibnamefont {Litzistorf}}, \bibinfo {author} {\bibfnamefont
  {P.}~\bibnamefont {Monbaron}}, \bibinfo {author} {\bibfnamefont
  {L.}~\bibnamefont {Monat}}, \bibinfo {author} {\bibfnamefont {J.-B.}\
  \bibnamefont {Page}}, \bibinfo {author} {\bibfnamefont {D.}~\bibnamefont
  {Perroud}}, \bibinfo {author} {\bibfnamefont {G.}~\bibnamefont {Ribordy}},
  \bibinfo {author} {\bibfnamefont {A.}~\bibnamefont {Rochas}}, \bibinfo
  {author} {\bibfnamefont {S.}~\bibnamefont {Robyr}}, \bibinfo {author}
  {\bibfnamefont {J.}~\bibnamefont {Tavares}}, \bibinfo {author} {\bibfnamefont
  {R.}~\bibnamefont {Thew}}, \bibinfo {author} {\bibfnamefont {P.}~\bibnamefont
  {Trinkler}}, \bibinfo {author} {\bibfnamefont {S.}~\bibnamefont {Ventura}},
  \bibinfo {author} {\bibfnamefont {R.}~\bibnamefont {Voirol}}, \bibinfo
  {author} {\bibfnamefont {N.}~\bibnamefont {Walenta}}, \ and\ \bibinfo
  {author} {\bibfnamefont {H.}~\bibnamefont {Zbinden}},\ }\href
  {http://stacks.iop.org/1367-2630/13/i=12/a=123001} {\bibfield  {journal}
  {\bibinfo  {journal} {New Journal of Physics}\ }\textbf {\bibinfo {volume}
  {13}},\ \bibinfo {pages} {123001} (\bibinfo {year} {2011})}\BibitemShut
  {NoStop}%
\bibitem [{\citenamefont {Ritter}\ \emph {et~al.}(2012)\citenamefont {Ritter},
  \citenamefont {Nolleke}, \citenamefont {Hahn}, \citenamefont {Reiserer},
  \citenamefont {Neuzner}, \citenamefont {Uphoff}, \citenamefont {Mucke},
  \citenamefont {Figueroa}, \citenamefont {Bochmann},\ and\ \citenamefont
  {Rempe}}]{ritter}%
  \BibitemOpen
  \bibfield  {author} {\bibinfo {author} {\bibfnamefont {S.}~\bibnamefont
  {Ritter}}, \bibinfo {author} {\bibfnamefont {C.}~\bibnamefont {Nolleke}},
  \bibinfo {author} {\bibfnamefont {C.}~\bibnamefont {Hahn}}, \bibinfo {author}
  {\bibfnamefont {A.}~\bibnamefont {Reiserer}}, \bibinfo {author}
  {\bibfnamefont {A.}~\bibnamefont {Neuzner}}, \bibinfo {author} {\bibfnamefont
  {M.}~\bibnamefont {Uphoff}}, \bibinfo {author} {\bibfnamefont
  {M.}~\bibnamefont {Mucke}}, \bibinfo {author} {\bibfnamefont
  {E.}~\bibnamefont {Figueroa}}, \bibinfo {author} {\bibfnamefont
  {J.}~\bibnamefont {Bochmann}}, \ and\ \bibinfo {author} {\bibfnamefont
  {G.}~\bibnamefont {Rempe}},\ }\href {\doibase 10.1038/nature11023} {\bibfield
   {journal} {\bibinfo  {journal} {Nature}\ }\textbf {\bibinfo {volume}
  {484}},\ \bibinfo {pages} {195} (\bibinfo {year} {2012})}\BibitemShut
  {NoStop}%
\bibitem [{\citenamefont {Garcia-Escartin}\ and\ \citenamefont
  {Chamorro-Posada}(2012)}]{garcia}%
  \BibitemOpen
  \bibfield  {author} {\bibinfo {author} {\bibfnamefont {J.~C.}\ \bibnamefont
  {Garcia-Escartin}}\ and\ \bibinfo {author} {\bibfnamefont {P.}~\bibnamefont
  {Chamorro-Posada}},\ }\href {\doibase 10.1103/PhysRevA.86.032334} {\bibfield
  {journal} {\bibinfo  {journal} {Phys. Rev. A}\ }\textbf {\bibinfo {volume}
  {86}},\ \bibinfo {pages} {032334} (\bibinfo {year} {2012})}\BibitemShut
  {NoStop}%
\bibitem [{\citenamefont {Chiorescu}\ \emph {et~al.}(2010)\citenamefont
  {Chiorescu}, \citenamefont {Groll}, \citenamefont {Bertaina}, \citenamefont
  {Mori},\ and\ \citenamefont {Miyashita}}]{chiorescu}%
  \BibitemOpen
  \bibfield  {author} {\bibinfo {author} {\bibfnamefont {I.}~\bibnamefont
  {Chiorescu}}, \bibinfo {author} {\bibfnamefont {N.}~\bibnamefont {Groll}},
  \bibinfo {author} {\bibfnamefont {S.}~\bibnamefont {Bertaina}}, \bibinfo
  {author} {\bibfnamefont {T.}~\bibnamefont {Mori}}, \ and\ \bibinfo {author}
  {\bibfnamefont {S.}~\bibnamefont {Miyashita}},\ }\href {\doibase
  10.1103/PhysRevB.82.024413} {\bibfield  {journal} {\bibinfo  {journal} {Phys.
  Rev. B}\ }\textbf {\bibinfo {volume} {82}},\ \bibinfo {pages} {024413}
  (\bibinfo {year} {2010})}\BibitemShut {NoStop}%
\bibitem [{\citenamefont {Wu}\ \emph {et~al.}(2010)\citenamefont {Wu},
  \citenamefont {George}, \citenamefont {Wesenberg}, \citenamefont {M\o{}lmer},
  \citenamefont {Schuster}, \citenamefont {Schoelkopf}, \citenamefont {Itoh},
  \citenamefont {Ardavan}, \citenamefont {Morton},\ and\ \citenamefont
  {Briggs}}]{wu}%
  \BibitemOpen
  \bibfield  {author} {\bibinfo {author} {\bibfnamefont {H.}~\bibnamefont
  {Wu}}, \bibinfo {author} {\bibfnamefont {R.~E.}\ \bibnamefont {George}},
  \bibinfo {author} {\bibfnamefont {J.~H.}\ \bibnamefont {Wesenberg}}, \bibinfo
  {author} {\bibfnamefont {K.}~\bibnamefont {M\o{}lmer}}, \bibinfo {author}
  {\bibfnamefont {D.~I.}\ \bibnamefont {Schuster}}, \bibinfo {author}
  {\bibfnamefont {R.~J.}\ \bibnamefont {Schoelkopf}}, \bibinfo {author}
  {\bibfnamefont {K.~M.}\ \bibnamefont {Itoh}}, \bibinfo {author}
  {\bibfnamefont {A.}~\bibnamefont {Ardavan}}, \bibinfo {author} {\bibfnamefont
  {J.~J.~L.}\ \bibnamefont {Morton}}, \ and\ \bibinfo {author} {\bibfnamefont
  {G.~A.~D.}\ \bibnamefont {Briggs}},\ }\href {\doibase
  10.1103/PhysRevLett.105.140503} {\bibfield  {journal} {\bibinfo  {journal}
  {Phys. Rev. Lett.}\ }\textbf {\bibinfo {volume} {105}},\ \bibinfo {pages}
  {140503} (\bibinfo {year} {2010})}\BibitemShut {NoStop}%
\bibitem [{\citenamefont {Lemr}\ \emph {et~al.}(2013)\citenamefont {Lemr},
  \citenamefont {Bartkiewicz}, \citenamefont {\ifmmode~\check{C}\else
  \v{C}\fi{}ernoch},\ and\ \citenamefont {Soubusta}}]{lemr}%
  \BibitemOpen
  \bibfield  {author} {\bibinfo {author} {\bibfnamefont {K.}~\bibnamefont
  {Lemr}}, \bibinfo {author} {\bibfnamefont {K.}~\bibnamefont {Bartkiewicz}},
  \bibinfo {author} {\bibfnamefont {A.}~\bibnamefont {\ifmmode~\check{C}\else
  \v{C}\fi{}ernoch}}, \ and\ \bibinfo {author} {\bibfnamefont {J.}~\bibnamefont
  {Soubusta}},\ }\href {\doibase 10.1103/PhysRevA.87.062333} {\bibfield
  {journal} {\bibinfo  {journal} {Phys. Rev. A}\ }\textbf {\bibinfo {volume}
  {87}},\ \bibinfo {pages} {062333} (\bibinfo {year} {2013})}\BibitemShut
  {NoStop}%
\bibitem [{\citenamefont {Lemr}\ and\ \citenamefont {\ifmmode~\check{C}\else
  \v{C}\fi{}ernoch}(2013)}]{lemr2}%
  \BibitemOpen
  \bibfield  {author} {\bibinfo {author} {\bibfnamefont {K.}~\bibnamefont
  {Lemr}}\ and\ \bibinfo {author} {\bibfnamefont {A.}~\bibnamefont
  {\ifmmode~\check{C}\else \v{C}\fi{}ernoch}},\ }\href {\doibase
  http://dx.doi.org/10.1016/j.optcom.2013.02.052} {\bibfield  {journal}
  {\bibinfo  {journal} {Optics Communications}\ }\textbf {\bibinfo {volume}
  {300}},\ \bibinfo {pages} {282 } (\bibinfo {year} {2013})}\BibitemShut
  {NoStop}%
\bibitem [{\citenamefont {Qu}\ \emph {et~al.}(2015)\citenamefont {Qu},
  \citenamefont {Zhou},\ and\ \citenamefont {Sheng}}]{qu}%
  \BibitemOpen
  \bibfield  {author} {\bibinfo {author} {\bibfnamefont {C.-C.}\ \bibnamefont
  {Qu}}, \bibinfo {author} {\bibfnamefont {L.}~\bibnamefont {Zhou}}, \ and\
  \bibinfo {author} {\bibfnamefont {Y.-B.}\ \bibnamefont {Sheng}},\ }\href
  {\doibase 10.1007/s10773-015-2539-9} {\bibfield  {journal} {\bibinfo
  {journal} {International Journal of Theoretical Physics}\ }\textbf {\bibinfo
  {volume} {54}},\ \bibinfo {pages} {3004} (\bibinfo {year}
  {2015})}\BibitemShut {NoStop}%
\bibitem [{\citenamefont {Chang}\ \emph {et~al.}(2007)\citenamefont {Chang},
  \citenamefont {S\o{}rensen}, \citenamefont {Demler},\ and\ \citenamefont
  {Lukin}}]{chang}%
  \BibitemOpen
  \bibfield  {author} {\bibinfo {author} {\bibfnamefont {D.~E.}\ \bibnamefont
  {Chang}}, \bibinfo {author} {\bibfnamefont {A.~S.}\ \bibnamefont
  {S\o{}rensen}}, \bibinfo {author} {\bibfnamefont {E.~A.}\ \bibnamefont
  {Demler}}, \ and\ \bibinfo {author} {\bibfnamefont {M.~D.}\ \bibnamefont
  {Lukin}},\ }\href {\doibase 10.1038/nphys708} {\bibfield  {journal} {\bibinfo
   {journal} {Nat Phys}\ }\textbf {\bibinfo {volume} {3}},\ \bibinfo {pages}
  {807} (\bibinfo {year} {2007})}\BibitemShut {NoStop}%
\bibitem [{\citenamefont {Neumeier}\ \emph {et~al.}(2013)\citenamefont
  {Neumeier}, \citenamefont {Leib},\ and\ \citenamefont {Hartmann}}]{neumeier}%
  \BibitemOpen
  \bibfield  {author} {\bibinfo {author} {\bibfnamefont {L.}~\bibnamefont
  {Neumeier}}, \bibinfo {author} {\bibfnamefont {M.}~\bibnamefont {Leib}}, \
  and\ \bibinfo {author} {\bibfnamefont {M.~J.}\ \bibnamefont {Hartmann}},\
  }\href {\doibase 10.1103/PhysRevLett.111.063601} {\bibfield  {journal}
  {\bibinfo  {journal} {Phys. Rev. Lett.}\ }\textbf {\bibinfo {volume} {111}},\
  \bibinfo {pages} {063601} (\bibinfo {year} {2013})}\BibitemShut {NoStop}%
\bibitem [{\citenamefont {Manzoni}\ \emph {et~al.}(2014)\citenamefont
  {Manzoni}, \citenamefont {Reiter}, \citenamefont {Taylor},\ and\
  \citenamefont {S\o{}rensen}}]{manzoni}%
  \BibitemOpen
  \bibfield  {author} {\bibinfo {author} {\bibfnamefont {M.~T.}\ \bibnamefont
  {Manzoni}}, \bibinfo {author} {\bibfnamefont {F.}~\bibnamefont {Reiter}},
  \bibinfo {author} {\bibfnamefont {J.~M.}\ \bibnamefont {Taylor}}, \ and\
  \bibinfo {author} {\bibfnamefont {A.~S.}\ \bibnamefont {S\o{}rensen}},\
  }\href {\doibase 10.1103/PhysRevB.89.180502} {\bibfield  {journal} {\bibinfo
  {journal} {Phys. Rev. B}\ }\textbf {\bibinfo {volume} {89}},\ \bibinfo
  {pages} {180502} (\bibinfo {year} {2014})}\BibitemShut {NoStop}%
\bibitem [{\citenamefont {Hoi}\ \emph {et~al.}(2011)\citenamefont {Hoi},
  \citenamefont {Wilson}, \citenamefont {Johansson}, \citenamefont {Palomaki},
  \citenamefont {Peropadre},\ and\ \citenamefont {Delsing}}]{hoi}%
  \BibitemOpen
  \bibfield  {author} {\bibinfo {author} {\bibfnamefont {I.-C.}\ \bibnamefont
  {Hoi}}, \bibinfo {author} {\bibfnamefont {C.~M.}\ \bibnamefont {Wilson}},
  \bibinfo {author} {\bibfnamefont {G.}~\bibnamefont {Johansson}}, \bibinfo
  {author} {\bibfnamefont {T.}~\bibnamefont {Palomaki}}, \bibinfo {author}
  {\bibfnamefont {B.}~\bibnamefont {Peropadre}}, \ and\ \bibinfo {author}
  {\bibfnamefont {P.}~\bibnamefont {Delsing}},\ }\href {\doibase
  10.1103/PhysRevLett.107.073601} {\bibfield  {journal} {\bibinfo  {journal}
  {Phys. Rev. Lett.}\ }\textbf {\bibinfo {volume} {107}},\ \bibinfo {pages}
  {073601} (\bibinfo {year} {2011})}\BibitemShut {NoStop}%
\bibitem [{\citenamefont {Agarwal}\ and\ \citenamefont
  {Huang}(2012)}]{agarwal}%
  \BibitemOpen
  \bibfield  {author} {\bibinfo {author} {\bibfnamefont {G.~S.}\ \bibnamefont
  {Agarwal}}\ and\ \bibinfo {author} {\bibfnamefont {S.}~\bibnamefont
  {Huang}},\ }\href {\doibase 10.1103/PhysRevA.85.021801} {\bibfield  {journal}
  {\bibinfo  {journal} {Phys. Rev. A}\ }\textbf {\bibinfo {volume} {85}},\
  \bibinfo {pages} {021801} (\bibinfo {year} {2012})}\BibitemShut {NoStop}%
\bibitem [{\citenamefont {Li}\ and\ \citenamefont {Zhu}(2012)}]{li}%
  \BibitemOpen
  \bibfield  {author} {\bibinfo {author} {\bibfnamefont {J.-J.}\ \bibnamefont
  {Li}}\ and\ \bibinfo {author} {\bibfnamefont {K.-D.}\ \bibnamefont {Zhu}},\
  }\href {\doibase http://dx.doi.org/10.1016/j.photonics.2012.05.001}
  {\bibfield  {journal} {\bibinfo  {journal} {Photonics and Nanostructures -
  Fundamentals and Applications}\ }\textbf {\bibinfo {volume} {10}},\ \bibinfo
  {pages} {553 } (\bibinfo {year} {2012})}\BibitemShut {NoStop}%
\bibitem [{\citenamefont {Lu}\ \emph {et~al.}(2014)\citenamefont {Lu},
  \citenamefont {Zhou}, \citenamefont {Kuang},\ and\ \citenamefont
  {Nori}}]{lu}%
  \BibitemOpen
  \bibfield  {author} {\bibinfo {author} {\bibfnamefont {J.}~\bibnamefont
  {Lu}}, \bibinfo {author} {\bibfnamefont {L.}~\bibnamefont {Zhou}}, \bibinfo
  {author} {\bibfnamefont {L.-M.}\ \bibnamefont {Kuang}}, \ and\ \bibinfo
  {author} {\bibfnamefont {F.}~\bibnamefont {Nori}},\ }\href {\doibase
  10.1103/PhysRevA.89.013805} {\bibfield  {journal} {\bibinfo  {journal} {Phys.
  Rev. A}\ }\textbf {\bibinfo {volume} {89}},\ \bibinfo {pages} {013805}
  (\bibinfo {year} {2014})}\BibitemShut {NoStop}%
\bibitem [{\citenamefont {Yan}\ and\ \citenamefont {Fan}(2014)}]{yan}%
  \BibitemOpen
  \bibfield  {author} {\bibinfo {author} {\bibfnamefont {W.-B.}\ \bibnamefont
  {Yan}}\ and\ \bibinfo {author} {\bibfnamefont {H.}~\bibnamefont {Fan}},\
  }\href {\doibase 10.1038/srep04820} {\bibfield  {journal} {\bibinfo
  {journal} {Sci Rep}\ }\textbf {\bibinfo {volume} {4}},\ \bibinfo {pages}
  {4820} (\bibinfo {year} {2014})}\BibitemShut {NoStop}%
\bibitem [{\citenamefont {Chirolli}\ \emph {et~al.}(2010)\citenamefont
  {Chirolli}, \citenamefont {Burkard}, \citenamefont {Kumar},\ and\
  \citenamefont {DiVincenzo}}]{prl104}%
  \BibitemOpen
  \bibfield  {author} {\bibinfo {author} {\bibfnamefont {L.}~\bibnamefont
  {Chirolli}}, \bibinfo {author} {\bibfnamefont {G.}~\bibnamefont {Burkard}},
  \bibinfo {author} {\bibfnamefont {S.}~\bibnamefont {Kumar}}, \ and\ \bibinfo
  {author} {\bibfnamefont {D.~P.}\ \bibnamefont {DiVincenzo}},\ }\href
  {\doibase 10.1103/PhysRevLett.104.230502} {\bibfield  {journal} {\bibinfo
  {journal} {Phys. Rev. Lett.}\ }\textbf {\bibinfo {volume} {104}},\ \bibinfo
  {pages} {230502} (\bibinfo {year} {2010})}\BibitemShut {NoStop}%
\bibitem [{\citenamefont {Hoffmann}\ \emph {et~al.}(2010)\citenamefont
  {Hoffmann}, \citenamefont {Deppe}, \citenamefont {Niemczyk}, \citenamefont
  {Wirth}, \citenamefont {Menzel}, \citenamefont {Wild}, \citenamefont {Huebl},
  \citenamefont {Mariantoni}, \citenamefont {Wei{\ss}l}, \citenamefont
  {Lukashenko}, \citenamefont {Zhuravel}, \citenamefont {Ustinov},
  \citenamefont {Marx},\ and\ \citenamefont {Gross}}]{hoffmann}%
  \BibitemOpen
  \bibfield  {author} {\bibinfo {author} {\bibfnamefont {E.}~\bibnamefont
  {Hoffmann}}, \bibinfo {author} {\bibfnamefont {F.}~\bibnamefont {Deppe}},
  \bibinfo {author} {\bibfnamefont {T.}~\bibnamefont {Niemczyk}}, \bibinfo
  {author} {\bibfnamefont {T.}~\bibnamefont {Wirth}}, \bibinfo {author}
  {\bibfnamefont {E.~P.}\ \bibnamefont {Menzel}}, \bibinfo {author}
  {\bibfnamefont {G.}~\bibnamefont {Wild}}, \bibinfo {author} {\bibfnamefont
  {H.}~\bibnamefont {Huebl}}, \bibinfo {author} {\bibfnamefont
  {M.}~\bibnamefont {Mariantoni}}, \bibinfo {author} {\bibfnamefont
  {T.}~\bibnamefont {Wei{\ss}l}}, \bibinfo {author} {\bibfnamefont
  {A.}~\bibnamefont {Lukashenko}}, \bibinfo {author} {\bibfnamefont {A.~P.}\
  \bibnamefont {Zhuravel}}, \bibinfo {author} {\bibfnamefont {A.~V.}\
  \bibnamefont {Ustinov}}, \bibinfo {author} {\bibfnamefont {A.}~\bibnamefont
  {Marx}}, \ and\ \bibinfo {author} {\bibfnamefont {R.}~\bibnamefont {Gross}},\
  }\href {\doibase http://dx.doi.org/10.1063/1.3522650} {\bibfield  {journal}
  {\bibinfo  {journal} {Appl. Phys. Lett.}\ }\textbf {\bibinfo {volume} {97}},\
  \bibinfo {eid} {222508} (\bibinfo {year} {2010})}\BibitemShut {NoStop}%
\bibitem [{\citenamefont {Giovannetti}\ \emph
  {et~al.}(2008{\natexlab{a}})\citenamefont {Giovannetti}, \citenamefont
  {Lloyd},\ and\ \citenamefont {Maccone}}]{pra78}%
  \BibitemOpen
  \bibfield  {author} {\bibinfo {author} {\bibfnamefont {V.}~\bibnamefont
  {Giovannetti}}, \bibinfo {author} {\bibfnamefont {S.}~\bibnamefont {Lloyd}},
  \ and\ \bibinfo {author} {\bibfnamefont {L.}~\bibnamefont {Maccone}},\ }\href
  {\doibase 10.1103/PhysRevA.78.052310} {\bibfield  {journal} {\bibinfo
  {journal} {Phys. Rev. A}\ }\textbf {\bibinfo {volume} {78}},\ \bibinfo
  {pages} {052310} (\bibinfo {year} {2008}{\natexlab{a}})}\BibitemShut
  {NoStop}%
\bibitem [{\citenamefont {Mariantoni}\ \emph {et~al.}(2008)\citenamefont
  {Mariantoni}, \citenamefont {Deppe}, \citenamefont {Marx}, \citenamefont
  {Gross}, \citenamefont {Wilhelm},\ and\ \citenamefont {Solano}}]{prb78}%
  \BibitemOpen
  \bibfield  {author} {\bibinfo {author} {\bibfnamefont {M.}~\bibnamefont
  {Mariantoni}}, \bibinfo {author} {\bibfnamefont {F.}~\bibnamefont {Deppe}},
  \bibinfo {author} {\bibfnamefont {A.}~\bibnamefont {Marx}}, \bibinfo {author}
  {\bibfnamefont {R.}~\bibnamefont {Gross}}, \bibinfo {author} {\bibfnamefont
  {F.~K.}\ \bibnamefont {Wilhelm}}, \ and\ \bibinfo {author} {\bibfnamefont
  {E.}~\bibnamefont {Solano}},\ }\href {\doibase 10.1103/PhysRevB.78.104508}
  {\bibfield  {journal} {\bibinfo  {journal} {Phys. Rev. B}\ }\textbf {\bibinfo
  {volume} {78}},\ \bibinfo {pages} {104508} (\bibinfo {year}
  {2008})}\BibitemShut {NoStop}%
\bibitem [{\citenamefont {Quintana}\ \emph {et~al.}(2013)\citenamefont
  {Quintana}, \citenamefont {Petersson}, \citenamefont {McFaul}, \citenamefont
  {Srinivasan}, \citenamefont {Houck},\ and\ \citenamefont {Petta}}]{prl110}%
  \BibitemOpen
  \bibfield  {author} {\bibinfo {author} {\bibfnamefont {C.~M.}\ \bibnamefont
  {Quintana}}, \bibinfo {author} {\bibfnamefont {K.~D.}\ \bibnamefont
  {Petersson}}, \bibinfo {author} {\bibfnamefont {L.~W.}\ \bibnamefont
  {McFaul}}, \bibinfo {author} {\bibfnamefont {S.~J.}\ \bibnamefont
  {Srinivasan}}, \bibinfo {author} {\bibfnamefont {A.~A.}\ \bibnamefont
  {Houck}}, \ and\ \bibinfo {author} {\bibfnamefont {J.~R.}\ \bibnamefont
  {Petta}},\ }\href {\doibase 10.1103/PhysRevLett.110.173603} {\bibfield
  {journal} {\bibinfo  {journal} {Phys. Rev. Lett.}\ }\textbf {\bibinfo
  {volume} {110}},\ \bibinfo {pages} {173603} (\bibinfo {year}
  {2013})}\BibitemShut {NoStop}%
\bibitem [{\citenamefont {Kyaw}\ \emph {et~al.}(2015)\citenamefont {Kyaw},
  \citenamefont {Felicetti}, \citenamefont {Romero}, \citenamefont {Solano},\
  and\ \citenamefont {Kwek}}]{kyaw}%
  \BibitemOpen
  \bibfield  {author} {\bibinfo {author} {\bibfnamefont {T.~H.}\ \bibnamefont
  {Kyaw}}, \bibinfo {author} {\bibfnamefont {S.}~\bibnamefont {Felicetti}},
  \bibinfo {author} {\bibfnamefont {G.}~\bibnamefont {Romero}}, \bibinfo
  {author} {\bibfnamefont {E.}~\bibnamefont {Solano}}, \ and\ \bibinfo {author}
  {\bibfnamefont {L.-C.}\ \bibnamefont {Kwek}},\ }\href
  {http://dx.doi.org/10.1038/srep08621} {\bibfield  {journal} {\bibinfo
  {journal} {Sci. Rep.}\ }\textbf {\bibinfo {volume} {5}},\ \bibinfo {pages}
  {8621} (\bibinfo {year} {2015})}\BibitemShut {NoStop}%
\bibitem [{\citenamefont {Hong}\ \emph {et~al.}(2012)\citenamefont {Hong},
  \citenamefont {Xiang}, \citenamefont {Zhu}, \citenamefont {Jiang},\ and\
  \citenamefont {Wu}}]{pra86}%
  \BibitemOpen
  \bibfield  {author} {\bibinfo {author} {\bibfnamefont {F.}~\bibnamefont
  {Hong}}, \bibinfo {author} {\bibfnamefont {Y.}~\bibnamefont {Xiang}},
  \bibinfo {author} {\bibfnamefont {Z.}~\bibnamefont {Zhu}}, \bibinfo {author}
  {\bibfnamefont {L.}~\bibnamefont {Jiang}}, \ and\ \bibinfo {author}
  {\bibfnamefont {L.}~\bibnamefont {Wu}},\ }\href {\doibase
  10.1103/PhysRevA.86.010306} {\bibfield  {journal} {\bibinfo  {journal} {Phys.
  Rev. A}\ }\textbf {\bibinfo {volume} {86}},\ \bibinfo {pages} {010306}
  (\bibinfo {year} {2012})}\BibitemShut {NoStop}%
\bibitem [{Note1()}]{Note1}%
  \BibitemOpen
  \bibinfo {note} {See Supplemental Material for derivations and additional
  analysys.}\BibitemShut {Stop}%
\bibitem [{\citenamefont {Devoret}(1995)}]{houches}%
  \BibitemOpen
  \bibfield  {author} {\bibinfo {author} {\bibfnamefont {M.~H.}\ \bibnamefont
  {Devoret}},\ }in\ \href@noop {} {\emph {\bibinfo {booktitle} {Quantum
  Fluctuations \'Ecole d'\'et\'e de Physique des Houches Session LXIII}}},\
  \bibinfo {series and number} {Les Houches},\ \bibinfo {editor} {edited by\
  \bibinfo {editor} {\bibfnamefont {S.}~\bibnamefont {Raimond}}, \bibinfo
  {editor} {\bibfnamefont {E.}~\bibnamefont {Giacobino}}, \ and\ \bibinfo
  {editor} {\bibfnamefont {J.}~\bibnamefont {Zinn-Justin}}}\ (\bibinfo
  {publisher} {Elsevier},\ \bibinfo {year} {1995})\ pp.\ \bibinfo {pages} {351
  -- 386}\BibitemShut {NoStop}%
\bibitem [{\citenamefont {Fan}\ \emph {et~al.}(2010)\citenamefont {Fan},
  \citenamefont {Kocaba\c{s}},\ and\ \citenamefont {Shen}}]{fan}%
  \BibitemOpen
  \bibfield  {author} {\bibinfo {author} {\bibfnamefont {S.}~\bibnamefont
  {Fan}}, \bibinfo {author} {\bibfnamefont {{\c{S}}.~E.}\ \bibnamefont
  {Kocaba\c{s}}}, \ and\ \bibinfo {author} {\bibfnamefont {J.-T.}\ \bibnamefont
  {Shen}},\ }\href {\doibase 10.1103/PhysRevA.82.063821} {\bibfield  {journal}
  {\bibinfo  {journal} {Phys. Rev. A}\ }\textbf {\bibinfo {volume} {82}},\
  \bibinfo {pages} {063821} (\bibinfo {year} {2010})}\BibitemShut {NoStop}%
\bibitem [{Note2()}]{Note2}%
  \BibitemOpen
  \bibinfo {note} {Given that the Lagrangian of the system is an even function
  (see supp. material), only odd levels of the transmon and SQUIDs are coupled
  to the transmission lines.}\BibitemShut {Stop}%
\bibitem [{\citenamefont {Koch}\ \emph {et~al.}(2007)\citenamefont {Koch},
  \citenamefont {Yu}, \citenamefont {Gambetta}, \citenamefont {Houck},
  \citenamefont {Schuster}, \citenamefont {Majer}, \citenamefont {Blais},
  \citenamefont {Devoret}, \citenamefont {Girvin},\ and\ \citenamefont
  {Schoelkopf}}]{koch}%
  \BibitemOpen
  \bibfield  {author} {\bibinfo {author} {\bibfnamefont {J.}~\bibnamefont
  {Koch}}, \bibinfo {author} {\bibfnamefont {T.~M.}\ \bibnamefont {Yu}},
  \bibinfo {author} {\bibfnamefont {J.}~\bibnamefont {Gambetta}}, \bibinfo
  {author} {\bibfnamefont {A.~A.}\ \bibnamefont {Houck}}, \bibinfo {author}
  {\bibfnamefont {D.~I.}\ \bibnamefont {Schuster}}, \bibinfo {author}
  {\bibfnamefont {J.}~\bibnamefont {Majer}}, \bibinfo {author} {\bibfnamefont
  {A.}~\bibnamefont {Blais}}, \bibinfo {author} {\bibfnamefont {M.~H.}\
  \bibnamefont {Devoret}}, \bibinfo {author} {\bibfnamefont {S.~M.}\
  \bibnamefont {Girvin}}, \ and\ \bibinfo {author} {\bibfnamefont {R.~J.}\
  \bibnamefont {Schoelkopf}},\ }\href {\doibase 10.1103/PhysRevA.76.042319}
  {\bibfield  {journal} {\bibinfo  {journal} {Phys. Rev. A}\ }\textbf {\bibinfo
  {volume} {76}},\ \bibinfo {pages} {042319} (\bibinfo {year}
  {2007})}\BibitemShut {NoStop}%
\bibitem [{Note3()}]{Note3}%
  \BibitemOpen
  \bibinfo {note} {Strictly speaking, after the addition of the nonlinear
  capacitor the behavior of the transmon will be modified. This does not affect
  our calculations, however, since they involve the behavior of the system as a
  whole and not that of the transmon by itself. For simplicity, we refer to
  ``transmons'' and ``SQUIDs'' (and their energy levels) throughout the
  paper.}\BibitemShut {Stop}%
\bibitem [{\citenamefont {Ilani}\ \emph {et~al.}(2006)\citenamefont {Ilani},
  \citenamefont {Donev}, \citenamefont {Kindermann},\ and\ \citenamefont
  {McEuen}}]{Ilani}%
  \BibitemOpen
  \bibfield  {author} {\bibinfo {author} {\bibfnamefont {S.}~\bibnamefont
  {Ilani}}, \bibinfo {author} {\bibfnamefont {L.~A.~K.}\ \bibnamefont {Donev}},
  \bibinfo {author} {\bibfnamefont {M.}~\bibnamefont {Kindermann}}, \ and\
  \bibinfo {author} {\bibfnamefont {P.~L.}\ \bibnamefont {McEuen}},\ }\href
  {\doibase 10.1038/nphys412} {\bibfield  {journal} {\bibinfo  {journal} {Nat.
  Phys.}\ }\textbf {\bibinfo {volume} {2}},\ \bibinfo {pages} {687} (\bibinfo
  {year} {2006})}\BibitemShut {NoStop}%
\bibitem [{Note4()}]{Note4}%
  \BibitemOpen
  \bibinfo {note} {Various systems give rise to nonlinear behavior of the
  capacitance as a function of the potential~\cite {Araujo,Gluskin,Colinge},
  but only with carbon nanotubes the energy spectrum of the capacitor has the
  form $E(V)= \protect \frac {C}{2} (V^2 + \alpha V^4)$~\cite {Akinwande2} for
  small voltages $V$, which is required for obtaining the interactions in
  Eq.~(\ref {eq:hc})}\BibitemShut {NoStop}%
\bibitem [{\citenamefont {Akinwande}\ \emph {et~al.}(2009)\citenamefont
  {Akinwande}, \citenamefont {Nishi},\ and\ \citenamefont {Wong}}]{Akinwande2}%
  \BibitemOpen
  \bibfield  {author} {\bibinfo {author} {\bibfnamefont {D.}~\bibnamefont
  {Akinwande}}, \bibinfo {author} {\bibfnamefont {Y.}~\bibnamefont {Nishi}}, \
  and\ \bibinfo {author} {\bibfnamefont {H.-S.}\ \bibnamefont {Wong}},\ }\href
  {\doibase 10.1109/TNANO.2008.2005185} {\bibfield  {journal} {\bibinfo
  {journal} {Nanotechnology, IEEE Transactions on}\ }\textbf {\bibinfo {volume}
  {8}},\ \bibinfo {pages} {31} (\bibinfo {year} {2009})}\BibitemShut {NoStop}%
\bibitem [{\citenamefont {Ithier}\ \emph {et~al.}(2005)\citenamefont {Ithier},
  \citenamefont {Collin}, \citenamefont {Joyez}, \citenamefont {Meeson},
  \citenamefont {Vion}, \citenamefont {Esteve}, \citenamefont {Chiarello},
  \citenamefont {Shnirman}, \citenamefont {Makhlin}, \citenamefont {Schriefl},\
  and\ \citenamefont {Sch\"on}}]{ithier}%
  \BibitemOpen
  \bibfield  {author} {\bibinfo {author} {\bibfnamefont {G.}~\bibnamefont
  {Ithier}}, \bibinfo {author} {\bibfnamefont {E.}~\bibnamefont {Collin}},
  \bibinfo {author} {\bibfnamefont {P.}~\bibnamefont {Joyez}}, \bibinfo
  {author} {\bibfnamefont {P.~J.}\ \bibnamefont {Meeson}}, \bibinfo {author}
  {\bibfnamefont {D.}~\bibnamefont {Vion}}, \bibinfo {author} {\bibfnamefont
  {D.}~\bibnamefont {Esteve}}, \bibinfo {author} {\bibfnamefont
  {F.}~\bibnamefont {Chiarello}}, \bibinfo {author} {\bibfnamefont
  {A.}~\bibnamefont {Shnirman}}, \bibinfo {author} {\bibfnamefont
  {Y.}~\bibnamefont {Makhlin}}, \bibinfo {author} {\bibfnamefont
  {J.}~\bibnamefont {Schriefl}}, \ and\ \bibinfo {author} {\bibfnamefont
  {G.}~\bibnamefont {Sch\"on}},\ }\href {\doibase 10.1103/PhysRevB.72.134519}
  {\bibfield  {journal} {\bibinfo  {journal} {Phys. Rev. B}\ }\textbf {\bibinfo
  {volume} {72}},\ \bibinfo {pages} {134519} (\bibinfo {year}
  {2005})}\BibitemShut {NoStop}%
\bibitem [{\citenamefont {Gardiner}\ and\ \citenamefont
  {Collett}(1985)}]{gardiner}%
  \BibitemOpen
  \bibfield  {author} {\bibinfo {author} {\bibfnamefont {C.~W.}\ \bibnamefont
  {Gardiner}}\ and\ \bibinfo {author} {\bibfnamefont {M.~J.}\ \bibnamefont
  {Collett}},\ }\href {\doibase 10.1103/PhysRevA.31.3761} {\bibfield  {journal}
  {\bibinfo  {journal} {Phys. Rev. A}\ }\textbf {\bibinfo {volume} {31}},\
  \bibinfo {pages} {3761} (\bibinfo {year} {1985})}\BibitemShut {NoStop}%
\bibitem [{\citenamefont {Giovannetti}\ \emph
  {et~al.}(2008{\natexlab{b}})\citenamefont {Giovannetti}, \citenamefont
  {Lloyd},\ and\ \citenamefont {Maccone}}]{Lloyd}%
  \BibitemOpen
  \bibfield  {author} {\bibinfo {author} {\bibfnamefont {V.}~\bibnamefont
  {Giovannetti}}, \bibinfo {author} {\bibfnamefont {S.}~\bibnamefont {Lloyd}},
  \ and\ \bibinfo {author} {\bibfnamefont {L.}~\bibnamefont {Maccone}},\ }\href
  {\doibase 10.1103/PhysRevLett.100.160501} {\bibfield  {journal} {\bibinfo
  {journal} {Phys. Rev. Lett.}\ }\textbf {\bibinfo {volume} {100}},\ \bibinfo
  {pages} {160501} (\bibinfo {year} {2008}{\natexlab{b}})}\BibitemShut
  {NoStop}%
\bibitem [{\citenamefont {de~Araujo}\ \emph {et~al.}(1995)\citenamefont
  {de~Araujo}, \citenamefont {Cuchiaro}, \citenamefont {McMillan},
  \citenamefont {Scott},\ and\ \citenamefont {Scott}}]{Araujo}%
  \BibitemOpen
  \bibfield  {author} {\bibinfo {author} {\bibfnamefont {C.~A.-P.}\
  \bibnamefont {de~Araujo}}, \bibinfo {author} {\bibfnamefont {J.~D.}\
  \bibnamefont {Cuchiaro}}, \bibinfo {author} {\bibfnamefont {L.~D.}\
  \bibnamefont {McMillan}}, \bibinfo {author} {\bibfnamefont {M.~C.}\
  \bibnamefont {Scott}}, \ and\ \bibinfo {author} {\bibfnamefont {J.~F.}\
  \bibnamefont {Scott}},\ }\href {\doibase 10.1038/374627a0} {\bibfield
  {journal} {\bibinfo  {journal} {Nature}\ }\textbf {\bibinfo {volume} {374}},\
  \bibinfo {pages} {627} (\bibinfo {year} {1995})}\BibitemShut {NoStop}%
\bibitem [{\citenamefont {Gluskin}(1985)}]{Gluskin}%
  \BibitemOpen
  \bibfield  {author} {\bibinfo {author} {\bibfnamefont {E.}~\bibnamefont
  {Gluskin}},\ }\href {\doibase 10.1080/00207218508939003} {\bibfield
  {journal} {\bibinfo  {journal} {International Journal of Electronics}\
  }\textbf {\bibinfo {volume} {58}},\ \bibinfo {pages} {63} (\bibinfo {year}
  {1985})}\BibitemShut {NoStop}%
\bibitem [{\citenamefont {Colinge}\ and\ \citenamefont {Colinge}()}]{Colinge}%
  \BibitemOpen
  \bibfield  {author} {\bibinfo {author} {\bibfnamefont {J.}~\bibnamefont
  {Colinge}}\ and\ \bibinfo {author} {\bibfnamefont {C.}~\bibnamefont
  {Colinge}},\ }\href@noop {} {\emph {\bibinfo {title} {Physics of
  Semiconductor Devices}}}\BibitemShut {NoStop}%
\end{thebibliography}

\begin{thebibliography}{25}%
\makeatletter
\providecommand \@ifxundefined [1]{%
 \@ifx{#1\undefined}
}%
\providecommand \@ifnum [1]{%
 \ifnum #1\expandafter \@firstoftwo
 \else \expandafter \@secondoftwo
 \fi
}%
\providecommand \@ifx [1]{%
 \ifx #1\expandafter \@firstoftwo
 \else \expandafter \@secondoftwo
 \fi
}%
\providecommand \natexlab [1]{#1}%
\providecommand \enquote  [1]{``#1''}%
\providecommand \bibnamefont  [1]{#1}%
\providecommand \bibfnamefont [1]{#1}%
\providecommand \citenamefont [1]{#1}%
\providecommand \href@noop [0]{\@secondoftwo}%
\providecommand \href [0]{\begingroup \@sanitize@url \@href}%
\providecommand \@href[1]{\@@startlink{#1}\@@href}%
\providecommand \@@href[1]{\endgroup#1\@@endlink}%
\providecommand \@sanitize@url [0]{\catcode `\\12\catcode `\$12\catcode
  `\&12\catcode `\#12\catcode `\^12\catcode `\_12\catcode `\%12\relax}%
\providecommand \@@startlink[1]{}%
\providecommand \@@endlink[0]{}%
\providecommand \url  [0]{\begingroup\@sanitize@url \@url }%
\providecommand \@url [1]{\endgroup\@href {#1}{\urlprefix }}%
\providecommand \urlprefix  [0]{URL }%
\providecommand \Eprint [0]{\href }%
\providecommand \doibase [0]{http://dx.doi.org/}%
\providecommand \selectlanguage [0]{\@gobble}%
\providecommand \bibinfo  [0]{\@secondoftwo}%
\providecommand \bibfield  [0]{\@secondoftwo}%
\providecommand \translation [1]{[#1]}%
\providecommand \BibitemOpen [0]{}%
\providecommand \bibitemStop [0]{}%
\providecommand \bibitemNoStop [0]{.\EOS\space}%
\providecommand \EOS [0]{\spacefactor3000\relax}%
\providecommand \BibitemShut  [1]{\csname bibitem#1\endcsname}%
\let\auto@bib@innerbib\@empty
\bibitem [{\citenamefont {DiCarlo}\ \emph {et~al.}(2009)\citenamefont
  {DiCarlo}, \citenamefont {Chow}, \citenamefont {Gambetta}, \citenamefont
  {Bishop}, \citenamefont {Johnson}, \citenamefont {Schuster}, \citenamefont
  {Majer}, \citenamefont {Blais}, \citenamefont {Frunzio}, \citenamefont
  {Girvin},\ and\ \citenamefont {Schoelkopf}}]{Sdicarlo}%
  \BibitemOpen
  \bibfield  {author} {\bibinfo {author} {\bibfnamefont {L.}~\bibnamefont
  {DiCarlo}}, \bibinfo {author} {\bibfnamefont {J.~M.}\ \bibnamefont {Chow}},
  \bibinfo {author} {\bibfnamefont {J.~M.}\ \bibnamefont {Gambetta}}, \bibinfo
  {author} {\bibfnamefont {L.~S.}\ \bibnamefont {Bishop}}, \bibinfo {author}
  {\bibfnamefont {B.~R.}\ \bibnamefont {Johnson}}, \bibinfo {author}
  {\bibfnamefont {D.~I.}\ \bibnamefont {Schuster}}, \bibinfo {author}
  {\bibfnamefont {J.}~\bibnamefont {Majer}}, \bibinfo {author} {\bibfnamefont
  {A.}~\bibnamefont {Blais}}, \bibinfo {author} {\bibfnamefont
  {L.}~\bibnamefont {Frunzio}}, \bibinfo {author} {\bibfnamefont {S.~M.}\
  \bibnamefont {Girvin}}, \ and\ \bibinfo {author} {\bibfnamefont {R.~J.}\
  \bibnamefont {Schoelkopf}},\ }\href {\doibase 10.1038/nature08121} {\bibfield
   {journal} {\bibinfo  {journal} {Nature}\ }\textbf {\bibinfo {volume}
  {460}},\ \bibinfo {pages} {240} (\bibinfo {year} {2009})}\BibitemShut
  {NoStop}%
\bibitem [{\citenamefont {Lucero}\ \emph {et~al.}(2012)\citenamefont {Lucero},
  \citenamefont {Barends}, \citenamefont {Chen}, \citenamefont {Kelly},
  \citenamefont {Mariantoni}, \citenamefont {Megrant}, \citenamefont
  {O'Malley}, \citenamefont {Sank}, \citenamefont {Vainsencher}, \citenamefont
  {Wenner}, \citenamefont {White}, \citenamefont {Yin}, \citenamefont
  {Cleland},\ and\ \citenamefont {Martinis}}]{Slucero}%
  \BibitemOpen
  \bibfield  {author} {\bibinfo {author} {\bibfnamefont {E.}~\bibnamefont
  {Lucero}}, \bibinfo {author} {\bibfnamefont {R.}~\bibnamefont {Barends}},
  \bibinfo {author} {\bibfnamefont {Y.}~\bibnamefont {Chen}}, \bibinfo {author}
  {\bibfnamefont {J.}~\bibnamefont {Kelly}}, \bibinfo {author} {\bibfnamefont
  {M.}~\bibnamefont {Mariantoni}}, \bibinfo {author} {\bibfnamefont
  {A.}~\bibnamefont {Megrant}}, \bibinfo {author} {\bibfnamefont
  {P.}~\bibnamefont {O'Malley}}, \bibinfo {author} {\bibfnamefont
  {D.}~\bibnamefont {Sank}}, \bibinfo {author} {\bibfnamefont {A.}~\bibnamefont
  {Vainsencher}}, \bibinfo {author} {\bibfnamefont {J.}~\bibnamefont {Wenner}},
  \bibinfo {author} {\bibfnamefont {T.}~\bibnamefont {White}}, \bibinfo
  {author} {\bibfnamefont {Y.}~\bibnamefont {Yin}}, \bibinfo {author}
  {\bibfnamefont {A.~N.}\ \bibnamefont {Cleland}}, \ and\ \bibinfo {author}
  {\bibfnamefont {J.~M.}\ \bibnamefont {Martinis}},\ }\href {\doibase
  10.1038/nphys2385} {\bibfield  {journal} {\bibinfo  {journal} {Nat. Phys.}\
  }\textbf {\bibinfo {volume} {8}},\ \bibinfo {pages} {719} (\bibinfo {year}
  {2012})}\BibitemShut {NoStop}%
\bibitem [{\citenamefont {Koch}\ \emph {et~al.}(2007)\citenamefont {Koch},
  \citenamefont {Yu}, \citenamefont {Gambetta}, \citenamefont {Houck},
  \citenamefont {Schuster}, \citenamefont {Majer}, \citenamefont {Blais},
  \citenamefont {Devoret}, \citenamefont {Girvin},\ and\ \citenamefont
  {Schoelkopf}}]{Skoch}%
  \BibitemOpen
  \bibfield  {author} {\bibinfo {author} {\bibfnamefont {J.}~\bibnamefont
  {Koch}}, \bibinfo {author} {\bibfnamefont {T.~M.}\ \bibnamefont {Yu}},
  \bibinfo {author} {\bibfnamefont {J.}~\bibnamefont {Gambetta}}, \bibinfo
  {author} {\bibfnamefont {A.~A.}\ \bibnamefont {Houck}}, \bibinfo {author}
  {\bibfnamefont {D.~I.}\ \bibnamefont {Schuster}}, \bibinfo {author}
  {\bibfnamefont {J.}~\bibnamefont {Majer}}, \bibinfo {author} {\bibfnamefont
  {A.}~\bibnamefont {Blais}}, \bibinfo {author} {\bibfnamefont {M.~H.}\
  \bibnamefont {Devoret}}, \bibinfo {author} {\bibfnamefont {S.~M.}\
  \bibnamefont {Girvin}}, \ and\ \bibinfo {author} {\bibfnamefont {R.~J.}\
  \bibnamefont {Schoelkopf}},\ }\href {\doibase 10.1103/PhysRevA.76.042319}
  {\bibfield  {journal} {\bibinfo  {journal} {Phys. Rev. A}\ }\textbf {\bibinfo
  {volume} {76}},\ \bibinfo {pages} {042319} (\bibinfo {year}
  {2007})}\BibitemShut {NoStop}%
\bibitem [{\citenamefont {de~Araujo}\ \emph {et~al.}(1995)\citenamefont
  {de~Araujo}, \citenamefont {Cuchiaro}, \citenamefont {McMillan},
  \citenamefont {Scott},\ and\ \citenamefont {Scott}}]{SAraujo}%
  \BibitemOpen
  \bibfield  {author} {\bibinfo {author} {\bibfnamefont {C.~A.-P.}\
  \bibnamefont {de~Araujo}}, \bibinfo {author} {\bibfnamefont {J.~D.}\
  \bibnamefont {Cuchiaro}}, \bibinfo {author} {\bibfnamefont {L.~D.}\
  \bibnamefont {McMillan}}, \bibinfo {author} {\bibfnamefont {M.~C.}\
  \bibnamefont {Scott}}, \ and\ \bibinfo {author} {\bibfnamefont {J.~F.}\
  \bibnamefont {Scott}},\ }\href {\doibase 10.1038/374627a0} {\bibfield
  {journal} {\bibinfo  {journal} {Nature}\ }\textbf {\bibinfo {volume} {374}},\
  \bibinfo {pages} {627} (\bibinfo {year} {1995})}\BibitemShut {NoStop}%
\bibitem [{\citenamefont {Gluskin}(1985)}]{SGluskin}%
  \BibitemOpen
  \bibfield  {author} {\bibinfo {author} {\bibfnamefont {E.}~\bibnamefont
  {Gluskin}},\ }\href {\doibase 10.1080/00207218508939003} {\bibfield
  {journal} {\bibinfo  {journal} {International Journal of Electronics}\
  }\textbf {\bibinfo {volume} {58}},\ \bibinfo {pages} {63} (\bibinfo {year}
  {1985})}\BibitemShut {NoStop}%
\bibitem [{\citenamefont {Sengouga}\ \emph {et~al.}(2015)\citenamefont
  {Sengouga}, \citenamefont {Boumaraf}, \citenamefont {Mari}, \citenamefont
  {Meftah}, \citenamefont {Jameel}, \citenamefont {Saqri}, \citenamefont
  {Azziz}, \citenamefont {Taylor},\ and\ \citenamefont {Henini}}]{Ssengouga}%
  \BibitemOpen
  \bibfield  {author} {\bibinfo {author} {\bibfnamefont {N.}~\bibnamefont
  {Sengouga}}, \bibinfo {author} {\bibfnamefont {R.}~\bibnamefont {Boumaraf}},
  \bibinfo {author} {\bibfnamefont {R.~H.}\ \bibnamefont {Mari}}, \bibinfo
  {author} {\bibfnamefont {A.}~\bibnamefont {Meftah}}, \bibinfo {author}
  {\bibfnamefont {D.}~\bibnamefont {Jameel}}, \bibinfo {author} {\bibfnamefont
  {N.~A.}\ \bibnamefont {Saqri}}, \bibinfo {author} {\bibfnamefont
  {M.}~\bibnamefont {Azziz}}, \bibinfo {author} {\bibfnamefont
  {D.}~\bibnamefont {Taylor}}, \ and\ \bibinfo {author} {\bibfnamefont
  {M.}~\bibnamefont {Henini}},\ }\href {\doibase
  http://dx.doi.org/10.1016/j.mssp.2015.03.043} {\bibfield  {journal} {\bibinfo
   {journal} {Materials Science in Semiconductor Processing}\ }\textbf
  {\bibinfo {volume} {36}},\ \bibinfo {pages} {156 } (\bibinfo {year}
  {2015})}\BibitemShut {NoStop}%
\bibitem [{\citenamefont {Colinge}\ and\ \citenamefont
  {Colinge}(2002)}]{SColinge}%
  \BibitemOpen
  \bibfield  {author} {\bibinfo {author} {\bibfnamefont {J.}~\bibnamefont
  {Colinge}}\ and\ \bibinfo {author} {\bibfnamefont {C.}~\bibnamefont
  {Colinge}},\ } {\emph
  {\bibinfo {title} {Physics of Semiconductor Devices}}}\ (\bibinfo
  {publisher} {Springer},\ \bibinfo {year} {2002})\BibitemShut {NoStop}%
\bibitem [{\citenamefont {Khan}\ \emph {et~al.}(2013)\citenamefont {Khan},
  \citenamefont {Lingalugari}, \citenamefont {Al-Amoody},\ and\ \citenamefont
  {Jain}}]{SKhan}%
  \BibitemOpen
  \bibfield  {author} {\bibinfo {author} {\bibfnamefont {J.}~\bibnamefont
  {Khan}}, \bibinfo {author} {\bibfnamefont {M.}~\bibnamefont {Lingalugari}},
  \bibinfo {author} {\bibfnamefont {F.}~\bibnamefont {Al-Amoody}}, \ and\
  \bibinfo {author} {\bibfnamefont {F.}~\bibnamefont {Jain}},\ }\href {\doibase
  10.1007/s11664-013-2713-x} {\bibfield  {journal} {\bibinfo  {journal}
  {Journal of Electronic Materials}\ }\textbf {\bibinfo {volume} {42}},\
  \bibinfo {pages} {3267} (\bibinfo {year} {2013})}\BibitemShut {NoStop}%
\bibitem [{\citenamefont {Ilani}\ \emph {et~al.}(2006)\citenamefont {Ilani},
  \citenamefont {Donev}, \citenamefont {Kindermann},\ and\ \citenamefont
  {McEuen}}]{SIlani}%
  \BibitemOpen
  \bibfield  {author} {\bibinfo {author} {\bibfnamefont {S.}~\bibnamefont
  {Ilani}}, \bibinfo {author} {\bibfnamefont {L.~A.~K.}\ \bibnamefont {Donev}},
  \bibinfo {author} {\bibfnamefont {M.}~\bibnamefont {Kindermann}}, \ and\
  \bibinfo {author} {\bibfnamefont {P.~L.}\ \bibnamefont {McEuen}},\ }\href
  {\doibase 10.1038/nphys412} {\bibfield  {journal} {\bibinfo  {journal} {Nat.
  Phys.}\ }\textbf {\bibinfo {volume} {2}},\ \bibinfo {pages} {687} (\bibinfo
  {year} {2006})}\BibitemShut {NoStop}%
\bibitem [{\citenamefont {Akinwande}\ \emph {et~al.}(2009)\citenamefont
  {Akinwande}, \citenamefont {Nishi},\ and\ \citenamefont {Wong}}]{SAkinwande2}%
  \BibitemOpen
  \bibfield  {author} {\bibinfo {author} {\bibfnamefont {D.}~\bibnamefont
  {Akinwande}}, \bibinfo {author} {\bibfnamefont {Y.}~\bibnamefont {Nishi}}, \
  and\ \bibinfo {author} {\bibfnamefont {H.-S.}\ \bibnamefont {Wong}},\ }\href
  {\doibase 10.1109/TNANO.2008.2005185} {\bibfield  {journal} {\bibinfo
  {journal} {Nanotechnology, IEEE Transactions on}\ }\textbf {\bibinfo {volume}
  {8}},\ \bibinfo {pages} {31} (\bibinfo {year} {2009})}\BibitemShut {NoStop}%
\bibitem [{Note5()}]{NoteS1}%
  \BibitemOpen
  \bibinfo {note} {If the energy is a function of the potential across the two
  plates of the capacitor whose minimum is not at $V=0$, a perturbation of the
  system will drive it to a region in the potential space of larger voltages,
  but in this region the analysis we make is not valid anymore because the
  dynamic variables that we quantize are small fluctuations of the
  potentials.}\BibitemShut {Stop}%
\bibitem [{\citenamefont {Bishop}(2010)}]{Sbishop}%
  \BibitemOpen
  \bibfield  {author} {\bibinfo {author} {\bibfnamefont {L.~S.}\ \bibnamefont
  {Bishop}},\ }\emph {\bibinfo {title} {Circuit Quantum Electrodynamics}},\
  \href {http://www.levbishop.org/thesis/} {Ph.D. thesis},\ \bibinfo  {school}
  {Yale University} (\bibinfo {year} {2010})\BibitemShut {NoStop}%
\bibitem [{\citenamefont {Devoret}(1995)}]{Shouches}%
  \BibitemOpen
  \bibfield  {author} {\bibinfo {author} {\bibfnamefont {M.~H.}\ \bibnamefont
  {Devoret}},\ }in\ \href@noop {} {\emph {\bibinfo {booktitle} {Quantum
  Fluctuations \'Ecole d'\'et\'e de Physique des Houches Session LXIII}}},\
  \bibinfo {series and number} {Les Houches},\ \bibinfo {editor} {edited by\
  \bibinfo {editor} {\bibfnamefont {S.}~\bibnamefont {Raimond}}, \bibinfo
  {editor} {\bibfnamefont {E.}~\bibnamefont {Giacobino}}, \ and\ \bibinfo
  {editor} {\bibfnamefont {J.}~\bibnamefont {Zinn-Justin}}}\ (\bibinfo
  {publisher} {Elsevier},\ \bibinfo {year} {1995})\ pp.\ \bibinfo {pages} {351
  -- 386}\BibitemShut {NoStop}%
\bibitem [{\citenamefont {Gardiner}\ and\ \citenamefont
  {Collett}(1985)}]{Sgardiner}%
  \BibitemOpen
  \bibfield  {author} {\bibinfo {author} {\bibfnamefont {C.~W.}\ \bibnamefont
  {Gardiner}}\ and\ \bibinfo {author} {\bibfnamefont {M.~J.}\ \bibnamefont
  {Collett}},\ }\href {\doibase 10.1103/PhysRevA.31.3761} {\bibfield  {journal}
  {\bibinfo  {journal} {Phys. Rev. A}\ }\textbf {\bibinfo {volume} {31}},\
  \bibinfo {pages} {3761} (\bibinfo {year} {1985})}\BibitemShut {NoStop}%
\bibitem [{\citenamefont {Neumeier}\ \emph {et~al.}(2013)\citenamefont
  {Neumeier}, \citenamefont {Leib},\ and\ \citenamefont {Hartmann}}]{Sneumeier}%
  \BibitemOpen
  \bibfield  {author} {\bibinfo {author} {\bibfnamefont {L.}~\bibnamefont
  {Neumeier}}, \bibinfo {author} {\bibfnamefont {M.}~\bibnamefont {Leib}}, \
  and\ \bibinfo {author} {\bibfnamefont {M.~J.}\ \bibnamefont {Hartmann}},\
  }\href {\doibase 10.1103/PhysRevLett.111.063601} {\bibfield  {journal}
  {\bibinfo  {journal} {Phys. Rev. Lett.}\ }\textbf {\bibinfo {volume} {111}},\
  \bibinfo {pages} {063601} (\bibinfo {year} {2013})}\BibitemShut {NoStop}%
\bibitem [{\citenamefont {Ithier}\ \emph {et~al.}(2005)\citenamefont {Ithier},
  \citenamefont {Collin}, \citenamefont {Joyez}, \citenamefont {Meeson},
  \citenamefont {Vion}, \citenamefont {Esteve}, \citenamefont {Chiarello},
  \citenamefont {Shnirman}, \citenamefont {Makhlin}, \citenamefont {Schriefl},\
  and\ \citenamefont {Sch\"on}}]{Sithier}%
  \BibitemOpen
  \bibfield  {author} {\bibinfo {author} {\bibfnamefont {G.}~\bibnamefont
  {Ithier}}, \bibinfo {author} {\bibfnamefont {E.}~\bibnamefont {Collin}},
  \bibinfo {author} {\bibfnamefont {P.}~\bibnamefont {Joyez}}, \bibinfo
  {author} {\bibfnamefont {P.~J.}\ \bibnamefont {Meeson}}, \bibinfo {author}
  {\bibfnamefont {D.}~\bibnamefont {Vion}}, \bibinfo {author} {\bibfnamefont
  {D.}~\bibnamefont {Esteve}}, \bibinfo {author} {\bibfnamefont
  {F.}~\bibnamefont {Chiarello}}, \bibinfo {author} {\bibfnamefont
  {A.}~\bibnamefont {Shnirman}}, \bibinfo {author} {\bibfnamefont
  {Y.}~\bibnamefont {Makhlin}}, \bibinfo {author} {\bibfnamefont
  {J.}~\bibnamefont {Schriefl}}, \ and\ \bibinfo {author} {\bibfnamefont
  {G.}~\bibnamefont {Sch\"on}},\ }\href {\doibase 10.1103/PhysRevB.72.134519}
  {\bibfield  {journal} {\bibinfo  {journal} {Phys. Rev. B}\ }\textbf {\bibinfo
  {volume} {72}},\ \bibinfo {pages} {134519} (\bibinfo {year}
  {2005})}\BibitemShut {NoStop}%
\bibitem [{\citenamefont {Fan}\ \emph {et~al.}(2010)\citenamefont {Fan},
  \citenamefont {Kocaba\c{s}},\ and\ \citenamefont {Shen}}]{Sfan}%
  \BibitemOpen
  \bibfield  {author} {\bibinfo {author} {\bibfnamefont {S.}~\bibnamefont
  {Fan}}, \bibinfo {author} {\bibfnamefont {{\c{S}}.~E.}\ \bibnamefont
  {Kocaba\c{s}}}, \ and\ \bibinfo {author} {\bibfnamefont {J.-T.}\ \bibnamefont
  {Shen}},\ }\href {\doibase 10.1103/PhysRevA.82.063821} {\bibfield  {journal}
  {\bibinfo  {journal} {Phys. Rev. A}\ }\textbf {\bibinfo {volume} {82}},\
  \bibinfo {pages} {063821} (\bibinfo {year} {2010})}\BibitemShut {NoStop}%
\bibitem [{\citenamefont {Reed}\ \emph {et~al.}(2012)\citenamefont {Reed},
  \citenamefont {DiCarlo}, \citenamefont {Nigg}, \citenamefont {Sun},
  \citenamefont {Frunzio}, \citenamefont {Girvin},\ and\ \citenamefont
  {Schoelkopf}}]{Sreed}%
  \BibitemOpen
  \bibfield  {author} {\bibinfo {author} {\bibfnamefont {M.~D.}\ \bibnamefont
  {Reed}}, \bibinfo {author} {\bibfnamefont {L.}~\bibnamefont {DiCarlo}},
  \bibinfo {author} {\bibfnamefont {S.~E.}\ \bibnamefont {Nigg}}, \bibinfo
  {author} {\bibfnamefont {L.}~\bibnamefont {Sun}}, \bibinfo {author}
  {\bibfnamefont {L.}~\bibnamefont {Frunzio}}, \bibinfo {author} {\bibfnamefont
  {S.~M.}\ \bibnamefont {Girvin}}, \ and\ \bibinfo {author} {\bibfnamefont
  {R.~J.}\ \bibnamefont {Schoelkopf}},\ }\href {\doibase 10.1038/nature10786}
  {\bibfield  {journal} {\bibinfo  {journal} {Nature}\ }\textbf {\bibinfo
  {volume} {482}},\ \bibinfo {pages} {382} (\bibinfo {year}
  {2012})}\BibitemShut {NoStop}%
\bibitem [{\citenamefont {Siewert}\ \emph {et~al.}(2006)\citenamefont
  {Siewert}, \citenamefont {Brandes},\ and\ \citenamefont {Falci}}]{SSiewert}%
  \BibitemOpen
  \bibfield  {author} {\bibinfo {author} {\bibfnamefont {J.}~\bibnamefont
  {Siewert}}, \bibinfo {author} {\bibfnamefont {T.}~\bibnamefont {Brandes}}, \
  and\ \bibinfo {author} {\bibfnamefont {G.}~\bibnamefont {Falci}},\ }\href
  {\doibase http://dx.doi.org/10.1016/j.optcom.2005.12.083} {\bibfield
  {journal} {\bibinfo  {journal} {Optics Communications}\ }\textbf {\bibinfo
  {volume} {264}},\ \bibinfo {pages} {435 } (\bibinfo {year} {2006})}
  \BibitemShut {NoStop}%
\bibitem [{\citenamefont {Feng}\ \emph {et~al.}(2010)\citenamefont {Feng},
  \citenamefont {Cai}, \citenamefont {Zhang}, \citenamefont {Fan},\ and\
  \citenamefont {Feng}}]{SFeng}%
  \BibitemOpen
  \bibfield  {author} {\bibinfo {author} {\bibfnamefont {Z.-B.}\ \bibnamefont
  {Feng}}, \bibinfo {author} {\bibfnamefont {Z.-L.}\ \bibnamefont {Cai}},
  \bibinfo {author} {\bibfnamefont {C.}~\bibnamefont {Zhang}}, \bibinfo
  {author} {\bibfnamefont {L.}~\bibnamefont {Fan}}, \ and\ \bibinfo {author}
  {\bibfnamefont {T.}~\bibnamefont {Feng}},\ }\href {\doibase
  http://dx.doi.org/10.1016/j.optcom.2010.01.002} {\bibfield  {journal}
  {\bibinfo  {journal} {Optics Communications}\ }\textbf {\bibinfo {volume}
  {283}},\ \bibinfo {pages} {1975 } (\bibinfo {year} {2010})}\BibitemShut
  {NoStop}%
\bibitem [{\citenamefont {Agarwal}\ and\ \citenamefont
  {Huang}(2012)}]{Sagarwal}%
  \BibitemOpen
  \bibfield  {author} {\bibinfo {author} {\bibfnamefont {G.~S.}\ \bibnamefont
  {Agarwal}}\ and\ \bibinfo {author} {\bibfnamefont {S.}~\bibnamefont
  {Huang}},\ }\href {\doibase 10.1103/PhysRevA.85.021801} {\bibfield  {journal}
  {\bibinfo  {journal} {Phys. Rev. A}\ }\textbf {\bibinfo {volume} {85}},\
  \bibinfo {pages} {021801} (\bibinfo {year} {2012})}\BibitemShut {NoStop}%
\bibitem [{\citenamefont {Hoi}\ \emph {et~al.}(2011)\citenamefont {Hoi},
  \citenamefont {Wilson}, \citenamefont {Johansson}, \citenamefont {Palomaki},
  \citenamefont {Peropadre},\ and\ \citenamefont {Delsing}}]{Shoi}%
  \BibitemOpen
  \bibfield  {author} {\bibinfo {author} {\bibfnamefont {I.-C.}\ \bibnamefont
  {Hoi}}, \bibinfo {author} {\bibfnamefont {C.~M.}\ \bibnamefont {Wilson}},
  \bibinfo {author} {\bibfnamefont {G.}~\bibnamefont {Johansson}}, \bibinfo
  {author} {\bibfnamefont {T.}~\bibnamefont {Palomaki}}, \bibinfo {author}
  {\bibfnamefont {B.}~\bibnamefont {Peropadre}}, \ and\ \bibinfo {author}
  {\bibfnamefont {P.}~\bibnamefont {Delsing}},\ }\href {\doibase
  10.1103/PhysRevLett.107.073601} {\bibfield  {journal} {\bibinfo  {journal}
  {Phys. Rev. Lett.}\ }\textbf {\bibinfo {volume} {107}},\ \bibinfo {pages}
  {073601} (\bibinfo {year} {2011})}\BibitemShut {NoStop}%
\bibitem [{\citenamefont {Lu}\ \emph {et~al.}(2014)\citenamefont {Lu},
  \citenamefont {Zhou}, \citenamefont {Kuang},\ and\ \citenamefont
  {Nori}}]{Slu}%
  \BibitemOpen
  \bibfield  {author} {\bibinfo {author} {\bibfnamefont {J.}~\bibnamefont
  {Lu}}, \bibinfo {author} {\bibfnamefont {L.}~\bibnamefont {Zhou}}, \bibinfo
  {author} {\bibfnamefont {L.-M.}\ \bibnamefont {Kuang}}, \ and\ \bibinfo
  {author} {\bibfnamefont {F.}~\bibnamefont {Nori}},\ }\href {\doibase
  10.1103/PhysRevA.89.013805} {\bibfield  {journal} {\bibinfo  {journal} {Phys.
  Rev. A}\ }\textbf {\bibinfo {volume} {89}},\ \bibinfo {pages} {013805}
  (\bibinfo {year} {2014})}\BibitemShut {NoStop}%
\bibitem [{\citenamefont {Lemr}\ \emph {et~al.}(2013)\citenamefont {Lemr},
  \citenamefont {Bartkiewicz}, \citenamefont {\ifmmode~\check{C}\else
  \v{C}\fi{}ernoch},\ and\ \citenamefont {Soubusta}}]{Slemr}%
  \BibitemOpen
  \bibfield  {author} {\bibinfo {author} {\bibfnamefont {K.}~\bibnamefont
  {Lemr}}, \bibinfo {author} {\bibfnamefont {K.}~\bibnamefont {Bartkiewicz}},
  \bibinfo {author} {\bibfnamefont {A.}~\bibnamefont {\ifmmode~\check{C}\else
  \v{C}\fi{}ernoch}}, \ and\ \bibinfo {author} {\bibfnamefont {J.}~\bibnamefont
  {Soubusta}},\ }\href {\doibase 10.1103/PhysRevA.87.062333} {\bibfield
  {journal} {\bibinfo  {journal} {Phys. Rev. A}\ }\textbf {\bibinfo {volume}
  {87}},\ \bibinfo {pages} {062333} (\bibinfo {year} {2013})}\BibitemShut
  {NoStop}%
\bibitem [{\citenamefont {Garcia-Escartin}\ and\ \citenamefont
  {Chamorro-Posada}(2012)}]{Sgarcia}%
  \BibitemOpen
  \bibfield  {author} {\bibinfo {author} {\bibfnamefont {J.~C.}\ \bibnamefont
  {Garcia-Escartin}}\ and\ \bibinfo {author} {\bibfnamefont {P.}~\bibnamefont
  {Chamorro-Posada}},\ }\href {\doibase 10.1103/PhysRevA.86.032334} {\bibfield
  {journal} {\bibinfo  {journal} {Phys. Rev. A}\ }\textbf {\bibinfo {volume}
  {86}},\ \bibinfo {pages} {032334} (\bibinfo {year} {2012})}\BibitemShut
  {NoStop}%
\end{thebibliography}
\end{document}